\documentclass[aps,prl,preprint,superscriptaddress]{revtex4}
\usepackage{times,xspace}
\usepackage{amsbsy,amssymb,amsmath,bm}
\usepackage{graphicx,color,epsfig,rotate}
\usepackage{amsfonts}
\begin{document}
\title{Radiation From Flux Flow In Josephson Junction Structures}
\author{L.N. Bulaevskii}
\affiliation{Los Alamos National Laboratory, Los Alamos, New Mexico 87545}
\author{A.E. Koshelev }
\affiliation{Materials Science Division, Argonne National Laboratory, Argonne, Illinois
60439 }
\date{\today}

\begin{abstract}
We derive the radiation power from a single Josephson junction (JJ)
and from layered superconductors in the flux-flow regime. For the JJ
case we formulate the boundary conditions for the electric and
magnetic fields at the edges of the superconducting leads using the
Maxwell equations in the dielectric media and find dynamic boundary
conditions for the phase difference in JJ which account for the
radiation. We derive the fraction of the power fed into JJ
transformed into the radiation. In a finite-length JJ this fraction
is determined by the dissipation inside JJ and it tends to unity as
dissipation vanishes independently of mismatch of the junction and
dielectric media impedances. We formulate also the dynamic boundary
conditions for the phase difference in intrinsic Josephson junctions
in highly anisotropic layered superconductors of the
Bi$_{2}$Sr$_{2}$CaCu$_{2}$O$_{8}$ type at the boundary with free
space. Using these boundary conditions, we solve equations for the
phase difference in the linear regime of Josephson oscillations for
rectangular and triangular lattices of Josephson vortices. In the
case of rectangular lattice for crystals with the thickness along
the $c$-axis much larger than the radiation wavelength we estimate
the radiation power per unit length in the direction of magnetic
field at the frequency 1 THz as $\sim N$ $\mu$W/cm for
Tl$_2$Ba$_2$CaCu$_2$O$_8$ and $\sim 0.04~N$ $\mu$W/cm for
Bi$_{2}$Sr$_{2}$CaCu$_{2}$O$_{8}$. For crystals with thickness
smaller than the radiation wavelength we found that the radiation
power in the resonance is independent on number of layers and can be
estimated at  1 THz as 0.5 W/cm (Tl$_2$Ba$_2$CaCu$_2$O$_8$) and 24
mW/cm (Bi$_{2}$Sr$_{2}$CaCu$_{2}$O$_{8}$). For rectangular lattice,
due to super-radiation regime,  up to half of power fed into the
crystal may be converted into the radiation. In the case of
triangular or random lattice in the direction perpendicular to the
layers the fraction of power converted into the radiation depends on
the dissipation rate and is much lower than for rectangular lattice
in the case of high-temperature superconductors with nodes in the
gap.
\end{abstract}

\maketitle


\section{Introduction}

In 1962 Josephson \cite{Jo} predicted electromagnetic radiation from
superconducting tunneling junction arguing that in the presence of voltage
$V$ across the junction the phase difference $\varphi$ changes with time
as $\partial\varphi/\partial t=2eV/\hbar$, while the tunneling current
density $J$ changes as $J=J_c\sin(2eVt/\hbar)$. Here $J_c$ is the critical
superconducting current via the junction. Thus the photons with the
Josephson frequency $\omega=2eV/\hbar$ may be emitted from Josephson
junction (JJ). Such a radiation from JJ into the waveguide with the power
$\mathcal{P}_{\mathrm{rad}}\sim10^{-12}$W has been detected by Dmitrenko
{\it et al.}\cite{Dm} and by Langenberg {\it et al.}
 \cite{Lan} soon after the Josephson prediction. The junction in
both experiments was subject to bias current and dc magnetic field applied
parallel to the junction. The dc magnetic field induces Josephson vortices
and in the presence of the transport current across the junction they move
to the JJ edge causing oscillations of the magnetic and electric fields
inside JJ and in superconducting leads around JJ. More precisely, flux
flow induces electromagnetic (Swihart) waves inside JJ, and they are
partially transmitted outside when hit JJ edges \cite{Sw}. From
measurements of the current-voltage ($I$-$V$) characteristics, Langenberg
{\emph et al.} \cite{Lan} found the power $\mathcal{P}=IV$ fed into the
junction and the fraction converted into radiation,
$Q=\mathcal{P}_{\mathrm{rad}} /\mathcal{P}\approx10^{-5}$. Theoretically
$Q$ was estimated as a ratio of the impedances of JJ modeled as a strip
line and the waveguide \cite{Lan,Kulik1}:
\begin{equation}
Q_{Z}=\frac{4Z_0Z_s}{(Z_0+Z_s)^2}\sim\left(\frac{8\lambda
d}{\varepsilon_{i}w^{2}}\right)^{1/2},\label{imp}
\end{equation}
where $Z_0$ is the waveguide impedance, $Z_s$ is the strip line
impedance in the limit $k_{\omega}w\ll 1$ and $Z_s\ll Z_0$. Here $\lambda$ is the London penetration length
of the superconducting leads, $d$ is the thickness of the insulating
layer, $\varepsilon_{i}$ is its dielectric constant, $k_{\omega}=\omega/c$ and $w$ is the
junction width. Eq.~(\ref{imp}) gives $\approx10^{-5}$ for the
studied junction in agreement with experimental value. Thus low
radiation power from JJ was attributed to the mismatch between the
impedances of the junction and the waveguide. To the best of our
knowledge, no deeper theoretical treatment of radiation was made
after that and the concept of the impedance mismatch for a single JJ
and later for layered superconductors with intrinsic JJ was
accepted by the community.

As radiation from the vortex flow in a single junction was found to
be quite low, a natural idea was proposed to use multiple lock-in
junctions. Then, in the super-radiation regime, radiation power
may be enhanced by the factor $N^2$, where
$N$ is the number of synchronized junctions inside the space of the radiation wavelength.
Intensive theoretical
and experimental study was done in this direction, see e.g.
Ref.~\onlinecite{Ust,Ped,Jain,Iv,Darula99,Kleiner,Ustinov}. However, an effective way to
synchronize many junctions was not found so far. There are two
problems to overcome for synchronization. The first one is the
technologically inevitable variation of parameters from junction to
junction, mainly $J_c$ parameter, which affects operating frequency
of the junction at given current. The second problem is that the
coherent locked-in flow of the Josephson vortices in different
junctions is unstable in a wide range of parameters.

The discovery of layered high-temperature superconductivity added new
breath into this activity. It was recognized that the Bi- and Tl-based
cuprate superconductors with weakly coupled superconducting CuO$_2$ layers
exhibit the same static and dynamical Josephson properties as artificial
tunneling junctions. In another words, the crystals
Bi$_{2}$Sr$_{2}$CaCu$_{2}$O$_{8}$ (BSCCO) and Tl$_2$Ba$_2$CaCu$_2$O$_8$
(TBCCO) represent a stack of many intrinsic Josephson junctions on the
atomic scale \cite{LD}. The first indication of the intrinsic Josephson
effect was observation of switching of individual junctions to the
resistive state in the $I$-$V$ characteristics \cite{Kl}. After that many
Josephson effects have been found in these systems including the
Fraunhofer patterns in the dependence of the critical current on the dc
magnetic field applied parallel to the layers \cite{Yur}, Josephson plasma
resonance \cite{Matsuda1} strongly affected by pancake vortices in the
presence of the magnetic field applied perpendicular to the layers
\cite{Oph,Matsuda2}, Shapiro steps in the $I$-$V$ characteristics induced
by external microwave radiation \cite{Wang01,YL}, and Fiske resonances
\cite{Irie98,Krasnov99,Kim04}

One can anticipate much smaller variations of intrinsic-junction
parameters in comparison with the artificially-fabricated junction arrays.
In addition, the intrinsic junctions are much closer to each other and one
can anticipate much stronger coupling between them. What is more, there
are many junctions on the scale of the radiation wavelength, and so they
will super-radiate when synchronized. Due to these advantages, the moving
Josephson vortex lattice in layered superconductors was proposed as a
source of monochromatic tunable continuous electromagnetic radiation in
the terahertz frequency range \cite{KoySSC95,Lat,Art,Tach,Nori}.
Experimentally, radiation from the high-temperature layered superconductor
BSCCO at relatively low frequencies, 7-16 GHz, was detected by
Hechtfischer {\emph et al.} \cite{hech}. Some indirect evidence of
radiation at higher frequencies has been recently reported by Kadowaki
\textit{et al.} \cite{Kad} and Batov {\it et al.} \cite{Batov} reported
radiation at the frequency 0.5 THz with power 1 pW from BSCCO mesa
consisting 100 junctions in zero dc magnetic field.

Therefore, it is interesting to estimate theoretically the possible
radiation power generated by a moving vortex lattice and find
optimal conditions for generation. For that we need to understand
the mechanism of conversion of the electromagnetic field associated with
the flux flow of the Josephson vortices into the electromagnetic
waves outside of the Josephson junction (JJ) and find limitations
imposed by this mechanism in Josephson tunneling structures. A
natural first step is to understand at the microscopic level the
conversion mechanism in a single JJ. Then such an approach may be
extended to layered structures.

In the first part we consider radiation from a single JJ. Then the method to treat radiation is extended
to intrinsic junctions in layered superconductors. We will show that super-radiation regime
is inherent to moving rectangular vortex lattice in such crystals. We will discuss the consequences of this
regime for the radiation power and $I$-$V$ characteristics.

\section{Radiation from a single Josephson junction}

As we discussed in the Introduction, a common way to evaluate the
radiation out of a single Josephson junction is to take product of the
total power supplied to the junction, $\mathcal{P}=IV$ and the impedance
mismatch coefficient $Q_Z$, Eq.~(\ref{imp}). A weak point of this
impedance approach for real, finite-length JJ is ignoring of multiple
reflections inside the junction, i.e., assumption that the propagating
electromagnetic wave has only one attempt to escape the JJ, decaying before
reflected wave reaches another edge. In fact, at low dissipation rate (low
temperature and low dielectric losses inside the insulating layer)
reflections lead to the formation of almost standing Swihart waves inside
JJ. In this case $Q$ \emph{strongly depends on the dissipation and
approaches unity as dissipation vanishes}. Then the question turns out to
be what are limitations on ${\cal P}$ and ${\cal P}_{{\rm rad}}$ rather
than on $Q$.

On the other hand, the standard analysis of \emph{transport properties} of
finite-size JJs uses sine-Gordon equation and zero-derivative boundary
conditions {\it for the oscillating part of the phase} at the edges
\cite{Kulik1,Kulik,Chang,Cirillo}. At the JJ edges vanishing oscillating
magnetic field $B=\partial\varphi/\partial x$ leads to zero Poynting
vector. Thus for such boundary conditions the outside radiation is absent
and all supplied power dissipates inside JJ.

In the following we reconsider this problem, discuss the power
conversion mechanism and derive the radiation power from a single JJ
into free space in the case when $w$ is much larger than the
wavelength of outcoming electromagnetic wave. This is not realistic limit,
but it is a simplest one which demonstrates the method to treat radiation and
which may be directly extended to the case of smaller $w$. Our rigorous
approach is based on solution of the Maxwell equations inside the
superconducting leads and in outside space, which allowed us to
formulate accurate dynamic boundary conditions for the oscillating
phase inside JJ. In the linear regime of Josephson oscillations
we obtained analytical results for
$\mathcal{P}_{\mathrm{rad}}$ and $Q$ using the perturbation theory.
In this regime we found that $Q\propto Q_{Z}\mathcal{N}$, where
$\mathcal{N}$ is the number of reflections before Swihart wave
decays inside JJ. At low temperatures in JJ made out of gapped
superconductors with perfect insulating layer $\mathcal{N}$ may be
large compensating small $Q_{Z}$. Our approach
also opens the way for numerical calculations in general case, when
linear approach is invalid.

The Josephson junctions with low level of the dissipation become available now
due to the perspective to use them as a qubits for quantum computing
\cite{Mart}. The radiation should be stronger in such junctions in comparison
with previously studied ones. This gives additional motivation to reconsider
the theoretical background of the radiation from JJ.

\begin{figure}[ptb]
\begin{center}
\includegraphics[width=0.4\textwidth,clip]{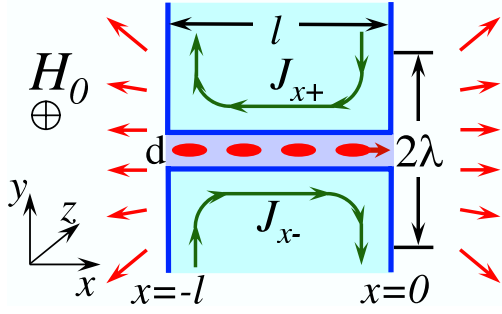}
\end{center}
\par
\vspace{-0.2in}\caption{A finite-size Josephson junction opened into
free space at both edges. Ellipses illustrate the moving vortex
lattice. Curved lines show the screening currents inside the
superconducting leads, arrows show radiation from area
$2\lambda$ around the junction. The applied dc magnetic field
$H_{0}$ is along the $z$-axis.} \label{Fig1}
\end{figure}

To find outside radiation due to the Josephson oscillations, one has to
match oscillating fields inside the junction and in the superconducting
leads with the wave solution outside the junction. Expressing the
oscillating fields inside the junctions via the phase difference, we
derive dynamic boundary conditions for the phase difference and relate the
Poynting vector of radiation to the phase difference at junction edge. The
final step is the solution of the equations for the phase difference
accounting for the dynamic boundary conditions and derivation of the
Poynting vector.

\subsection{ Equation for the phase difference}

We consider a JJ with the length $l\gg\lambda$ located at $-l<x<0$ and
bounded by dielectric with dielectric constant $\varepsilon_{d}$, see Fig.
\ref{Fig1}. Strength of the coupling in the junction is characterized by
the Josephson current density $J_{c}$ and related parameters: the
Josephson length, $\lambda_{J}=\sqrt{c\Phi_{0} /(8\pi^{2}\Lambda J_{c})}$,
and plasma frequency,$\ \omega_{p}=\sqrt{8\pi^{2} dcJ_{c}/(\varepsilon_{i}
\Phi_{0})}$, where $\Lambda=2\lambda+d$. We consider the simplest
situation when the JJ width along $z$ direction, $w$, is larger than both
$\lambda_{J}$ and wavelength of outcoming electromagnetic wave. We will
consider straight Josephson vortices along the external magnetic field. In
this approach the problem becomes one-dimensional for the phase difference
and two-dimensional for the electric and magnetic fields in the outer
space as both of them become $z$-independent. We consider junction in
resistive state and assume that the junction phase, $\bar{\varphi}$,
oscillates with the Josephson frequency $\omega$ generating oscillating
electric ($E_{x}$ and $E_{y}$) and magnetic ($B_{z}$) fields inside the
junction and in the superconducting leads. Our task is to find spatial
distribution of these fields and match them with outside fields to find
equation and boundary conditions for $\tilde{\varphi}$.

We first derive equation for the oscillating magnetic field inside the
superconducting leads. We use complex representation for the oscillating
fields and phase, e.g.,
\[
\bar{\varphi}(x,t)= \langle\bar{\varphi}(x,t)\rangle_t+
\sum_{\omega}\operatorname{Re}[\bar{\varphi}_{\omega}(x)e^{-i\omega t}].
\]
Here $\langle \ldots \rangle_{t}$ means time average. The phase gradient
is connected by the following relation with the magnetic field inside the
junction and supercurrents flowing along the junction at the opposite
sides
\begin{equation}
\nabla_{x}\bar{\varphi}=\frac{8\pi^{2}\lambda^{2}}{c\Phi_{0} }\left[  J_{x+}
-J_{x-}\right]  -\frac{2\pi d}{\Phi_{0}}B_{z}.\label{eB}
\end{equation}
From Maxwell equations, material equation for current inside superconductor,
$\mathbf{J}=\left(  c/4\pi\lambda^{2}\right)  \left[  \left(  \Phi_{0}
/2\pi\right)  \nabla\phi-\mathbf{A}\right]  +\sigma_{q}\mathbf{E}$, London
relation for the electric field, $\mathbf{E}=-\left(  4\pi\lambda^{2}
i\omega/c\right)  \mathbf{J}$, and Eq. (\ref{eB}), we derive the following
equation for the oscillating magnetic field inside the leads ($-l<x<0$) at
$d\ll\lambda$
\begin{eqnarray}
&&\left(  \nabla_{x}^{2}+\nabla_{y}^{2}\right)  B_{z}(\omega)-\frac{B_{z}
(\omega)} {\lambda_{\omega}^{2}}=\frac{\Phi_{0}}{2\pi\lambda^{2}}
\frac{\partial\bar{\varphi}_{\omega} }{\partial x}\delta(y),\label{main} \\
&&\lambda_{\omega}^{-2}=\lambda^{-2}-\varepsilon_{s}
k_{\omega}^{2}+4\pi ik_{\omega}\sigma_{q}/c, \ \ \ k_{\omega}=\omega/c. \label{ls}
\end{eqnarray}
We ignore small contribution to the magnetic
field from the dc current flowing via the junction. The ac electric
field inside the superconducting leads is
$E_{y}(\omega,x,y)=i\lambda_{\omega} ^{2}k_{\omega}\nabla_{x}B_{z}
(\omega,x,y)$. To obtain the total electric field in the junction
area, one has to account also for the field inside the dielectric
layer. This gives
\begin{equation}
E_{y}(\omega,x,y)\!=\!-\frac{i\omega\Phi_{0}}{2\pi
c}\bar{\varphi}_{\omega
}(x)\delta(y)\!+\!i\lambda_{\omega}^{2}k_{\omega}\nabla_{x}B_{z}(\omega
,x,y).\label{Ey_total}
\end{equation}
The boundary condition for $\bar{\varphi}_{\omega}(x)$ follows from the
boundary condition for the electric field
\begin{equation}
-i\omega\lbrack\varepsilon_{d}E_{x}(+0,y)-\varepsilon E_{x}(-0,y)]=4\pi
J_{x}(-0,y).\nonumber
\end{equation}
As $E_{x}(-0,y)=-i\omega(4\pi\lambda^{2}/c^{2})J_{x}(-0,y)$, we obtain
\begin{equation}
4\pi(1-\varepsilon k_{0}^{2}\lambda^{2})J_{x}(-0,y)=-i\omega\varepsilon_{d}
E_{x}(+0,y).
\end{equation}
As the electric field $E_{x}(+0,y)$ is continuous at $y=0$, this means that
$J_{x}(0,y)$ also must be continuous and from Eq.~(\ref{eB}) in the limit
$d\ll\lambda$ we obtain
\begin{equation}
\nabla_{x}\bar{\varphi}_{\omega}(0)=\nabla_{x}\bar{\varphi}_{\omega
}(-l)=0.\label{zerobc}
\end{equation}
We can represent solution of Eq. (\ref{main}) near the edge $x=0$ as
\begin{align}
&  B_{z}(x,y)=B_{b}(x,y)-\frac{\Phi_{0}}{(2\pi\lambda)^{2}}\int_{-\infty} ^{0}
\nabla_{x^{\prime}}\bar{\varphi}_{\omega}(x^{\prime})dx^{\prime}
\times\label{MagField}\\
& [K_0(c_-)-K_0(c_+)], \ \ c_{\pm}=\sqrt{(x\pm x^{\prime})^{2}+y^{2}}/\lambda_{\omega},
\nonumber
\end{align}
where $K_{0}\left(  z\right)  $ is the modified Bessel function and
$B_{b}(x,y) $ is the solution of the homogeneous equation
\begin{equation}
\left(  \nabla_{x}^{2}+\nabla_{y}^{2}\right)  B_{b}-\lambda_{\omega}
^{-2}B_{b} =0,
\label{Beq}\end{equation}
with the boundary condition $B_{b}(-0,y) =B_{z}(+0,y)$. As a
function of $x$, $B_{b}(x,y)$ decays at distance $\sim\lambda$ from
the boundary. For its Fourier transform along the $y$ direction, we
obtain [$\kappa_{x} =(\lambda_{\omega}^{-2}+k_{y}^{2})^{1/2}$]
\[
B_{b}(\omega,x,k_y)=B_{z}(+0,k_{y})\exp(-\kappa_{x}|x|).
\]
For static case, distribution of the magnetic field near the edge of JJ
has been recently derived by A.\ Gurevich (unpublished).

Using Maxwell equation $\nabla_{x}B_z=\left(  i\omega\varepsilon_{i}/c\right)
E_{y}-\left(  4\pi/c\right)  J_{y}$, Josephson relation $E_{y} (x,0)=-i\omega
\Phi_{0}\bar{\varphi}_{\omega}(x)/(2\pi cd)$, and Eq.\ (\ref{MagField}), we
find equation for the oscillating phase
\begin{align}
&  \left(
\frac{\omega^{2}}{\omega_{p}^{2}}+\alpha_{t}\frac{i\omega}
{\omega_{p}}\right)
\bar{\varphi}_{\omega}(x)+\frac{\lambda_{J}^{2}} {\pi\lambda}
\int_{-\infty}^{0} dx^{\prime}\nabla_{x}\left[  K_{0}\left(
\frac{x-x^{\prime}} {\lambda_{\omega} }\right)  + \right. \nonumber\\
&  \left.  K_{0}\left(  \frac{x+x^{\prime}}{\lambda_{\omega}}\right)  \right]
\nabla_{x^{\prime}}\bar{\varphi}_{\omega}(x^{\prime}) - \frac{c\nabla_{x}B_{b}
(x,0)}{4\pi J_{c}}=s_{\omega}(x),\label{PhaseEq}
\end{align}
where $s_{\omega}(x)$ is the complex amplitude of the oscillating Josephson
current, $\sin\left[  \bar{\varphi}(x,t)\right]  =\operatorname{Re}[s_{\omega
}(x)\exp(-i\omega t)]$, and $\alpha_{t}=\omega_{p}\sigma_{t}\Phi_{0}/(2\pi
J_{c}cd)$ is damping due to the tunneling quasiparticle conductivity.
Similar nonlocal equation has been derived by Gurevich \cite{Gur}.
In the case $\lambda\ll\lambda_{J}$ and $|x|\gg\lambda$ non-locality is not
essential and Eq.~(\ref{PhaseEq}) can be reduced to usual local approximate
equation
\begin{equation}
\left(
\frac{\omega^{2}}{\omega_{p}^{2}}\!+\!\alpha_{t}\frac{i\omega}
{\omega_{p} }\right) \varphi_{\omega}\!+\!\lambda_{J}^{2}\left(
1\!-\!\alpha_{q}\frac{i\omega }{\omega_{p}}\right)
\nabla_{x}^{2}\varphi_{\omega}\!=\!s_{\omega}
(x),\label{localPhaseEq}
\end{equation}
where $\alpha_{q}=2\pi\lambda^{2}\omega_{p}\sigma_{q}/c^{2}$ is the
dissipation due to quasiparticles inside superconductor and we used expansion
$\lambda_{\omega}/\lambda\approx1-2\pi i\lambda^{2}k_{\omega}\sigma_{q}/c$
neglecting $k_{\omega}^{2}$ term.
\subsection{Boundary conditions and the Poynting vector}
Near the boundary situation is more complicated because, in addition to
smoothly changing part, the phase has component decaying at distances of
order $\lambda$ from the boundary. This extra phase is smaller than the
smooth phase by the factor $\sim\lambda/\lambda_{J}$ but it has comparable
derivative. Our purpose is to derive accurate boundary condition for the
smooth phase, ${\varphi}_{\omega}(x)$, obeying Eq. (\ref{localPhaseEq}).
For this we integrate Eq. (\ref{PhaseEq}) from intermediate distance
$-x_{i}$ with $\lambda\ll x_{i}\ll\lambda_{J}$ up to the boundary $x=0$.
Neglecting small terms proportional to $x_{i}$ and $B_{b}(-x_{i},0)$ [for
$B_{z}(0,y)\ll \Phi_{0}/(4\pi\lambda^{2})$], using local approximation at
$x=-x_{i}$, we obtain the boundary condition for the smooth phase in a
very simple form
\begin{equation}
\nabla_{x}\varphi_{\omega}(0)\approx\nabla_{x}\bar{\varphi}_{\omega}
(-x_{i})=-\frac{\lambda}{\lambda_{\omega}}\frac{4\pi\lambda}{\Phi_{0}}
B_{z}(\omega,\mathbf{r}\! =\!0).\label{BoundGradSmPhase}
\end{equation}
This equation can be compared with condition (\ref{zerobc}) for \emph{the
total phase}. Eq.~(\ref{BoundGradSmPhase}) allows us reduce the problem to
solution of local equation (\ref{localPhaseEq}) for smooth phase with modified
boundary conditions and avoid solving exact integral equation for the total
phase, Eqs.~(\ref{zerobc}) and (\ref{PhaseEq}).

For that we need to express $B_{z}(\mathbf{r}\! =\!0)$ via ${\varphi}_{\omega
}$. Relation between the electric and magnetic field at the boundary is determined
by properties of outside media.

\subsection{Outside dielectric media}

We consider first outside dielectric media with dielectric constant
$\varepsilon_{d}$. In the situation $wk_{\omega}\gg1$, in the
straight-vortices approach, outside fields are $z$-independent. In
addition, if we assume that the dielectric media is infinite in $y$
direction and the thickness of the leads is much larger than $\lambda$,
we can use Fourier transform in this direction. In this
case the Fourier components of fields with
$|k_{y}|<\sqrt{\varepsilon_{d}}k_{\omega}$ propagate in the media
while the field components with
$|k_{y}|>\sqrt{\varepsilon_{d}}k_{\omega}$ decay. In particular, for
$E_{y}(\omega,x,k_{y})$ we have
\begin{equation}
E_{y}(\omega,x,k_{y})=
\genfrac{\{}{.}{0pt}{}{E_{y}(\omega,0,k_{y})\exp[i\sqrt{\varepsilon
_{d}k_{\omega}^{2}-k_{y}^{2}}\mathrm{sign}(\omega)x], \ \ \ \text{ for
}|k_{y}
|<\sqrt{\varepsilon_{d}}k_{\omega},}{E_{y}(\omega,0,k_{y})\exp[-\sqrt
{k_{y}^{2}-\varepsilon_{d}k_{\omega}^{2}}x], \ \ \ \text{ for
}|k_{y}|>\sqrt {\varepsilon_{d}}\left\vert k_{\omega}\right\vert .}
\label{Ey_ky}
\end{equation}
Other field components, $E_{x}$ and $B_{z}$, can be expressed via
$E_{y}(\omega,0,k_{y})$. First, $E_{x}(\omega,x,k_{y})$ can be
obtained from Eq.\ (\ref{Ey_ky}) and Maxwell equation
$\nabla~\mathbf{E}=0$ and then $B_{z}(\omega,x,k_{y})$ can be
obtained using the Maxwell equation
$(\nabla\times\mathbf{E})_{z}=ik_{\omega}B_{z}$ which gives
\begin{eqnarray}
&&B_{z}(\omega,x,k_{y}) =\frac{\varepsilon_{d}|k_{\omega}|}{\sqrt
{\varepsilon_{d}k_{\omega}^{2}-k_{y}^{2}}}E_{y}(\omega,0,k_{y})\exp
[i\sqrt{\varepsilon_{d}k_{\omega}^{2}-k_{y}^{2}}\mathrm{sign}(\omega)x], \ \ \ \text{
for }|k_{y}|<\sqrt{\varepsilon_{d}}k_{\omega},\nonumber\\
&&B_{z}(\omega,x,k_{y})
=\frac{-i\varepsilon_{d}k_{\omega}}{\sqrt{k_{y}^{2}-\varepsilon_{d}
k_{\omega}^{2}}}E_{y}(\omega,0,k_{y})\exp[-\sqrt{k_{y}^{2}-\varepsilon
_{d}k_{\omega}^{2}}x], \ \ \ \text{ for
}|k_{y}|>\sqrt{\varepsilon_{d}}k_{\omega },\label{Bz_ky}
\end{eqnarray}
In particular, this gives relation between fields at the boundary,
which we will use to formulate the boundary conditions for the phase
\begin{align}
&  B_{z}(0,k_{y})=\zeta(\omega,k_{y})E_{y}(0,k_{y}),\label{BoutEout}\\
&  \zeta(\omega,k_{y})=
\genfrac{\{}{.}{0pt}{}{|k_{\omega}|\varepsilon_{d}/\sqrt{\varepsilon
_{d}k_{\omega}^{2}-k_{y}^{2}}\text{, for
}|k_{y}|<\sqrt{\varepsilon_{d}
}|k_{\omega}|,}{-ik_{\omega}\varepsilon_{d}/\sqrt{k_{y}^{2}-\varepsilon
_{d}k_{\omega}^{2}}\text{, for
}|k_{y}|>\sqrt{\varepsilon_{d}}\left\vert k_{\omega}\right\vert .}
\nonumber
\end{align}
Note again that the term $\zeta(\omega,k_{y})$ for $|k_{y}|<\sqrt
{\varepsilon_{d}}k_{\omega}$ originates from outcoming
electromagnetic wave (radiation), while the term
$\zeta(\omega,k_{y})$ for $|k_{y}|>\sqrt
{\varepsilon_{d}}k_{\omega}$ is due to the wave decaying at distance
$\sim(k_{y}^{2}-\varepsilon_{d}k_{\omega}^{2})^{-1/2}$ from the lead
boundaries. The latter term does not carry energy out of the
junction. For completeness, we also present this important relation
in the frequency-space representation
\begin{align}
B_{z}(\omega,0,y)  & =\int dy^{\prime}U(\omega,y-y^{\prime})E_{y}
(\omega,0,y^{\prime}),\label{tot}\\
U(\omega,y)  & =\frac{\varepsilon_{d}}{2}\left[  |k_{\omega}|J_{0}
(\sqrt{\varepsilon_{d}}k_{\omega}y)+ik_{\omega}N_{0}(\sqrt{\varepsilon_{d}
}k_{\omega}y)\right]  .\nonumber
\end{align}
where $J_{0}(z)$ and $N_{0}(z)$ are the Bessel functions, and in the
time-space representation
\begin{equation}
B_{z}(t,0,y)=\sqrt{\varepsilon_{d}}\frac{\partial}{\partial t}\int
dt^{\prime }dy^{\prime}\frac{\Theta\lbrack
c_{d}^{2}(t-t^{\prime})^{2}-(y-y^{\prime}
)^{2}]}{\sqrt{c_{d}^{2}(t-t^{\prime})^{2}-(y-y^{\prime})^{2}}}E_{y}(t^{\prime
},0,y^{\prime})
\label{space-time}\end{equation}
where $\Theta(x)=1$ if $x>0$ and $0$ if $x<0$ and $c_{d}=c/\sqrt
{\varepsilon_{d}}$.

Now we will relate boundary fields with the phase. From Eqs.
(\ref{Ey_total}) and (\ref{MagField}) follows the relation
\begin{align}
&  E_{y}(0,k_{y})=i\lambda_{\omega}^{2}k_{\omega}\kappa_{x}B_{z}
(0,k_{y})-\label{Eout-s}\\
&  \frac{ik_{\omega}\Phi_{0}}{2\pi}\left(  \varphi_{\omega}(0)-\frac
{\lambda_{\omega}^{2}}{\lambda^{2}}\int_{-\infty}^{0}\exp\left(
\kappa _{x}x\right)  \nabla_{x}\varphi_{\omega}(x)dx\right)
.\nonumber
\end{align}
This relation and Eq.\ (\ref{BoutEout}) allow us to express the
boundary fields via the phase distribution
\begin{align}
&  E_{y}(0,k_{y})=\frac{B_{z}(0,k_{y})}{\zeta(\omega,k_{y})},\ \ \
B_{z} (0,k_{y})=\frac{\Phi_{0}}{2\pi}\frac{i\zeta
k_{\omega}}{1-i\zeta
\lambda_{\omega}^{2}k_{\omega}\kappa_{x}}\times\nonumber\\
&  \left(
-\varphi_{\omega}(0)+\frac{\lambda_{\omega}^{2}}{\lambda^{2}}
\int_{-\infty}^{0}\exp\left(  \kappa_{x}x\right)
\nabla_{x}\varphi_{\omega }(x)dx\right)  .\label{Bout}
\end{align}
Typically, $\zeta\lambda_{\omega}^{2}k_{\omega}\kappa_{x}\sim\lambda
\omega/c\ll1$. Also
$\nabla_{x}\varphi_{\omega}(x)\sim\varphi_{\omega }(x)/\lambda_{J}$
and the integral term in Eq.~(\ref{Bout}) is smaller than
$\varphi_{\omega}(0)$ by the parameter $\lambda/\lambda_{J}$.
Therefore, we obtain
\begin{equation}
E_{y}(0,k_{y})\!\approx\!-\frac{\Phi_{0}}{2\pi}ik_{\omega}\varphi_{\omega
}(0),\ B_{z}(0,k_{y})\!\approx\!-\frac{\Phi_{0}}{2\pi}i\zeta
k_{\omega} \varphi_{\omega}(0).
\end{equation}
Within this approximation, the magnetic field at the junction edge
which determines the boundary condition (\ref{BoundGradSmPhase}) is
\begin{align}
&  B_{z}(\mathbf{r}\!=\!0)\approx-\frac{\Phi_{0}}{2\pi}ik_{\omega}
\varphi_{\omega}(0)Z(\omega),\ \ \ Z(\omega)\approx\label{bz}\\
& \int_{-\pi/d}^{\pi/d}\frac{dk_{y}}{2\pi}\zeta(\omega,k_{y})\approx
\frac{\varepsilon_{d}|k_{\omega}|}{2}-i\frac{\varepsilon_{d}k_{\omega}}{\pi
}\ln\frac{\pi/d}{\sqrt{\varepsilon_{d}}|k_{\omega}|}.\nonumber
\end{align}
Finally, we obtain the boundary conditions for smooth oscillating
phase at both edges, $x=0,-l$, in a finite-length JJ:
\begin{equation}
\nabla_{x}{\varphi}_{\omega}(x)\approx\pm2i\lambda
k_{\omega}Z(\omega ){\varphi}_{\omega}(x),\text{for
}x=0,-l.\label{BoundSmoothFinal}
\end{equation}
Radiation outside JJ to the right is given by the Poynting vector,
$\mathcal{P}_{\mathrm{rad}}\!=\!(c/4\pi)\!\int\!dy\langle
E_{y}(0,y,t)B_{z} (0,y,t)\rangle_{t}$,
\begin{equation}
\mathcal{P}_{\mathrm{rad}}\approx\frac{\varepsilon_{d}w\omega^{3}\Phi_{0}^{2}
}{64\pi^{3}c^{2}}|\varphi_{\omega}(0)|^{2}.\label{Poynting}
\end{equation}
We also present the phase equation and boundary conditions in the
time representation. Introducing the dimensionless variables
$\tau=\omega_{p}t$ and $u=x/\lambda_{J}$ and using Eq.\
(\ref{localPhaseEq}), we write the equation for the total smooth
phase $\varphi(\tau,u)$, which includes both oscillating and
non-oscillating components
\begin{equation}
\left[  \frac{\partial^{2}}{\partial\tau^{2}}-\nabla_{u}^{2}+\left(
\alpha_{t}-\alpha_{q}\nabla_{u}^{2}\right)
\frac{\partial}{\partial\tau }\right]
\varphi+\sin\varphi=0.\label{st}
\end{equation}
Using Eqs.~(\ref{BoundGradSmPhase}), (\ref{bz}), and
(\ref{BoundSmoothFinal}) and adding the boundary condition for the
static phase, we obtain the dynamic boundary condition:
\begin{align}
&  \nabla_{u}\varphi(\tau,0)=-b-\int_{-\infty}^{\tau}d\tau^{\prime}
\mathcal{K}(\tau-\tau^{\prime})\frac{\partial\varphi(\tau^{\prime}
,0)}{\partial\tau^{\prime}},\label{t}\\
&
\mathcal{K}(\tau)=\frac{d\varepsilon_{d}}{2\pi\varepsilon_{i}\lambda_{J}
}\frac{\partial}{\partial\tau}\left[
\frac{\mathcal{F}(\xi\tau)}{\tau }\right]  ,\ \
\mathcal{F}(v)=\int_{0}^{v}dzJ_{0}(z).\nonumber
\end{align}
Here, $\xi=\pi c/(\omega_{p}d\sqrt{\varepsilon_{d}})$, $J_{0}(z)$ is
the Bessel function, $b=4\pi\lambda\lambda_{J}H_{0}/\Phi_{0}$ with
$\mathbf{H} _{0}\parallel z$ being the applied dc magnetic field,
$b\ll\lambda_{J} /\lambda$. For the edge $x=-l$ we need to change
sign of $\mathcal{K}$. Due to the non-analytical $\omega$-dependence
of $Z(\omega)$, the kernel $\mathcal{K}(\tau)$ is irregular at
$d\rightarrow0$ and then $\tau \rightarrow0$. Singular frequency
dependence of $Z(\omega)$ is due to retardation caused by
electromagnetic wave propagation inside dielectric, Eq.~(\ref{space-time}).
Thus boundary conditions also exhibit retardation effect.

\subsection{Outside superconducting screen}

\begin{figure}
\begin{center}
 \epsfig{file=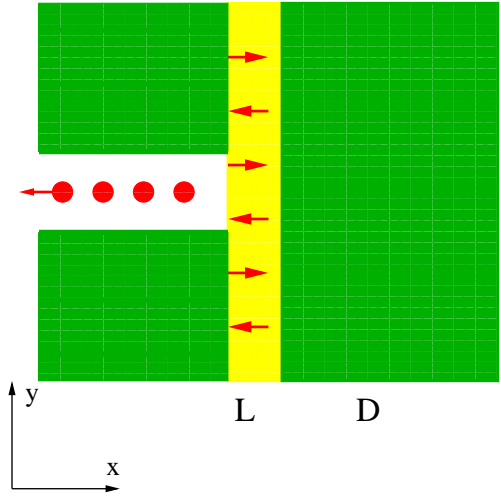,width=0.4\textwidth,angle=0}
\end{center}
\caption{ The Josephson junctions at $x<0$, dielectric at $0<x<L\ll
k_{\omega}^{-1}$ and  superconducting screen with the London penetration
length $\lambda_s$ and the thickness $D\gg \lambda_s$. Such a screen
effectively reflects the Swihart wave back into the junction. It introduces additional
interaction of vortices in different junctions in the case of multiple
parallel junctions. \label{Fig2} }
\end{figure}

Let us consider now a junction with superconducting screen at the right
side separated from the layered crystal by the dielectric with the thickness
$L$ small in comparison with the wave length of the electromagnetic wave
in this dielectric, see Fig.~\ref{Fig2}. Such a screen prevents the
radiation from the right side increasing it from other side. In the case
of many parallel junctions the screen enhances their interaction as was
suggested by Ivanchenko \cite{Iv}.

We consider the case $k_{\omega}L\ll 1$ and $L\ll \lambda_s$.
Fields inside the superconducting
screen obey Eq.~(\ref{Beq}). The solution is
\begin{eqnarray}
&&E_y(\omega,x,y)=
\frac{1}{(2\pi)^2}\int_{-\infty}^{+\infty}dk_yG(k_y)e^{-\kappa_xx+ik_yy},
 \ \ \kappa_x=\sqrt{\lambda_{s,\omega}^{-2}+k_y^2}, \nonumber\\
&&E_x(\omega,x,y)=
-\frac{1}{(2\pi)^2}\int_{-\infty}^{+\infty}dk_y
G(k_y)\frac{ik_y}{\kappa_x}e^{-\kappa_xx+ik_yy}, \label{scr}\\
&&B_z(\omega,x,y)=
-\frac{1}{(2\pi)^2}\int_{-\infty}^{+\infty}dk_y
G(k_y)\frac{\lambda_{s,\omega}^{-2}}{ik_{\omega}\kappa_x}e^{-\kappa_xx+ik_yy},
\nonumber
\end{eqnarray}
where $\lambda_{s,\omega}$, given by Eq.~(\ref{ls}), characterizes the superconducting screen and
$\lambda_s$ is the London penetration length of the screen. From this
solution we obtain
\begin{eqnarray}
&&B_z(\omega,{\bf r}=0)=
-\frac{\Phi_0}{2\pi}ik_{\omega}\varphi_{\omega}(0)Z_s(\omega), \\
&&Z_s(\omega)=-\frac{i}{\pi k_{\omega}\lambda_{s,\omega}^2}
\ln\left(\frac{2\lambda_s}{L}\right). \nonumber
\end{eqnarray}
This leads to the boundary condition
\begin{equation}
\frac{\partial\varphi_{\omega}(0)}{\partial u}=\eta\varphi_{\omega}(0),
\ \ \ \eta\approx \frac{2\lambda_J\lambda}{\pi\lambda_s^2}\ln\left(\frac{2\lambda_s}{L}\right).
\end{equation}
Note that $\eta\gg 1$. This results in the condition that $\varphi_{\omega}(0)$ is
very close to zero for screened edge of the junction, while at the open
edge the condition is that space derivative is close to zero. There is flow of energy
from closed edge proportional to the small parameter $4\pi \lambda_s^2\sigma_q/c$
due to quasiparticle dissipation inside the screen.

\subsection{Solution for the phase difference, $I$-$V$ characteristics,
radiation and dissipation power}

Now we solve analytically Eq.~(\ref{st}) using the perturbation theory
with respect to the Josephson current \cite{Kulik,Cirillo} in the limit
$|b-\tilde{\omega}|b\gg1$. Taking solution as
\begin{equation}
\varphi(\tau,u)=\tilde{\omega}\tau-bu+\theta(\tau,u),\ \ \ \theta(\tau
,u)\ll1,\label{pert}
\end{equation}
and expanding $\sin[\varphi(\tau,u)]$, we see that $\theta_{\omega}
(u)\equiv\theta(\tilde{\omega},u)$ obeys reduced versions of equation
(\ref{localPhaseEq}) with $s_{\omega}=-e^{ibu}/i$,
\begin{equation}
\left[  \nabla_{u}^{2}+\tilde{\omega}^{2}+i\tilde{\omega}(\alpha_{t}
-\alpha_{s}\nabla_{u}^{2})\right]  \theta_{\omega}(u)=-e^{ibu}/i.
\end{equation}
and boundary conditions (\ref{BoundSmoothFinal})
\begin{equation}
\frac{d\varphi}{du}=\pm i\tilde{\omega}\beta\varphi,\ \text{for\ }
u=0,-\tilde{l}\label{BoundCondRed}
\end{equation}
with
\[
\beta=\beta_{0}\left(  |\tilde{\omega}|-\frac{2i\tilde{\omega}}{\pi}\ln
\frac{\pi c/d}{\sqrt{\varepsilon_{d}}\omega_{p}|\tilde{\omega}|}\right)
,\ \beta _{0}=\frac{\varepsilon_{d}d}{2\varepsilon_{i}\lambda_{J}}
\]
Solution for $\theta(\tilde{\omega},u)$ given by,
\begin{equation}
\theta_{\omega}(u)\approx-\frac{e^{ibu}/i}{\tilde{\omega}^{2}-b^{2}
+i\alpha_{b}\tilde{\omega}}+a_{1}e^{ip_{\omega}u}+a_{2}e^{-ip_{\omega}
u},\label{aa}
\end{equation}
describes the moving vortex lattice (the first term) and reflected Swihart
waves propagating to the right and left. Here $\alpha_{b}=\alpha_{t}
+\alpha_{q}b^{2}$ and $p_{\omega}\!=\sqrt{\left(  \tilde{\omega}^{2}
+i\alpha_{t}\tilde{\omega}\right)  /\left(  1-i\alpha_{q}\tilde{\omega
}\right)  }\approx\!\tilde{\omega}\!+\!i\alpha/2$ with $\alpha\!=\!\alpha
_{t}\!+\!\alpha_{q}\tilde{\omega}^{2}$. For a JJ with both open edges
finding $a_{1}$ and $a_{2}$ from the boundary conditions
(\ref{BoundCondRed}) we obtain the phase distribution
\begin{align*}
& \theta _{\omega }(u)=-\frac{\exp \left( ibu\right) /i}{\tilde{\omega}
^{2}-b^{2}+i\alpha _{b}\tilde{\omega}}- \\
& \frac{\left( b\!-\!\beta \tilde{\omega}\right) \left[ \cos [p_{\omega }(
\tilde{l}\!+\!u)]\!-\!i\tilde{\beta}\sin [p_{\omega }(\tilde{l}\!+\!u)]
\right] \!-\!\left( b\!+\!\beta \tilde{\omega}\right) \exp
(\!-ib\tilde{l}) \left[ \cos \left( p_{\omega }u\right)
\!+\!i\tilde{\beta}\sin (p_{\omega }u) \right] }{\left(
\tilde{\omega}^{2}-b^{2}+i\alpha _{b}\tilde{\omega}\right) p_{\omega
}\left[ \left( 1+\tilde{\beta}^{2}\right) \sin (p_{\omega }\tilde{l
})+2i\tilde{\beta}\cos (p_{\omega }\tilde{l})\right] }
\end{align*}
with $\tilde{\beta}\equiv \beta \tilde{\omega}/p_{\omega }$ and
$\tilde{l}=l/\lambda_{J}$. Alternatively, the phase was obtained by
expansion with respect to eigenmodes \cite {Kulik,Cirillo}. In particular,
the boundary values which determine outside irradiation are given by
\begin{align*}
\theta _{\omega }(0)& =\frac{i}{\tilde{\omega}^{2}-b^{2}+i\alpha
_{b}\tilde{ \omega}}\left[ 1-\frac{\left( b-\beta \tilde{\omega}\right)
\left[ \cos (p_{\omega }\tilde{l})-i\tilde{\beta}\sin (p_{\omega
}\tilde{l})\right] -\left( b+\beta \tilde{\omega}\right) \exp \left(
-ib\tilde{l}\right) }{ ip_{\omega }\left[ \left(
1+\tilde{\beta}^{2}\right) \sin (p_{\omega}
\tilde{l})+2i\tilde{\beta}\cos (p_{\omega }\tilde{l})\right] }\right]  \\
\theta _{\omega }(-\tilde{l})& =\frac{i\exp \left( -ib\tilde{l}\right) }{
\tilde{\omega}^{2}-b^{2}+i\alpha _{b}\tilde{\omega}}\left[ 1-\frac{\left(
b-\beta \tilde{\omega}\right) \exp \left( ib\tilde{l}\right) -\left( b+
\tilde{\beta}\tilde{\omega}\right) \left[ \cos \left( p_{\omega }\tilde{l}
\right) -i\beta \sin \left( p_{\omega }\tilde{l}\right) \right]
}{ip_{\omega }\left[ \left( 1+\tilde{\beta}^{2}\right) \sin (p_{\omega
}\tilde{l})+2i \tilde{\beta}\cos (p_{\omega }\tilde{l})\right] }\right]
\end{align*}
These cumbersome formulas can be significantly simplified in the regime
$b\gg\tilde{\omega}$, weak dissipation $\alpha_{t}$, $\alpha_{q}\ll1$, and
large impedance mismatch, $|\beta| \ll 1 $. In this case, keeping only
Fiske resonance terms, we obtain
\begin{align}
&
\theta_{\omega}(0)\approx\lbrack\cos(p_{\omega}\tilde{l})-\exp(-ib\tilde
{l})]/(\tilde{\omega}bD), \nonumber\\
&  \theta_{\omega}(-l)\approx\lbrack1-\exp(-ib\tilde{l})\cos(p_{\omega}
\tilde{l})]/(\tilde{\omega}bD),\label{solsing}\\
&  D\approx\sin(\tilde{\omega}\tilde{l})+i(2\beta+\alpha\tilde{l}/2)\cos
(\tilde{\omega}\tilde{l}),\nonumber
\end{align}
In these approximate results we also neglected the term $\operatorname{Im}
\left[  \beta(\omega)\right]  $, which only slightly shifts resonance
positions. Radiation to the right and left, $\mathcal{P}_{\mathrm{rad}}
^{r,l}(\tilde{\omega},b)$, is determined by the values $|\theta_{\omega
}(0)|^{2}$ and $|\theta_{\omega}(-l)|^{2}$. At low dissipation, $\alpha
\tilde{l}\ll1$, we derive
\begin{align}
&  \mathcal{P}_{\mathrm{rad}}^{r,l}\approx\frac{\Phi_{0}^{2}\omega\omega
_{p}^{2}w}{64\pi^{3}c^{2}b^{2}}\frac{1\!-\!2\cos(b\tilde{l})\cos(\tilde
{\omega}\tilde{l})\!+\!\cos^{2}(\tilde{\omega}\tilde{l})\!\pm\!\rho}{|D|^{2}
},\label{genepower}\\
&  \rho=2(\alpha\tilde{l}+2\beta)[1-\cos(b\tilde{l})\cos(\omega\tilde
{l})]/b.\nonumber
\end{align}
The radiation reaches maxima for almost standing Swihart waves at
frequencies $\tilde{\omega}\!=\!\tilde{\omega}_{n}\!=\!\pi n/\tilde{l}$
with $n\!=\!1,2,\ldots$ when
$\cos(b\tilde{l})\cos(\tilde{\omega}_{n}\tilde {l})\!\neq\!1$. The
resonance width is determined by both, the dissipation, $\alpha\tilde{l}$,
and by the radiation, $\beta$. The perturbation theory is valid in
resonance for $|\theta_{\omega}(0)|\sim(b\alpha\tilde{l}\tilde
{\omega})^{-1}<1$ and the radiation power in this linear regime is quite
small, $\mathcal{P}_{\mathrm{rad}}/w\lesssim10^{-6}$ and $\lesssim1$
$\mu$W/cm at $\nu\!=\!10$ GHz and 1 THz, respectively. The asymmetry of
radiation described by $\rho$ is small.

Next we derive the dc current at voltage $V=\Phi_{0}\omega/(2\pi c)$ and
estimate $Q$. The current $I$ via JJ is given by the tunneling
quasiparticle contribution, $I_{t}=\sigma_{t}Vlw/d$, and the Josephson
current contribution, $I_{s}$,
\begin{align}
I_{s}(\omega) &  \approx J_{c}lwi_{J}\label{c}\\
i_{J} &
=\tilde{l}^{-1}\int_{-\tilde{l}}^{0}du\langle\cos(\tilde{\omega}\tau-bu){\theta
}(\tau,u)\rangle_{\tau}.
\end{align}
The exact result for the reduced Josephson current, $i_J$, is given by
\begin{align}
i_{J} &  =\frac{\alpha_{b}\tilde{\omega}/2}{\left(  \tilde{\omega}^{2}
-b^{2}\right)  ^{2}+\alpha_{b}^{2}\tilde{\omega}^{2}}\label{JosCurrExact}\\
&  +\operatorname{Im}\left[  \frac{\left( b^{2}+\tilde{\beta}^{2}p_{\omega
}^{2}\right)  \left[  \cos( b\tilde{l}) \!-\!\cos(  p_{\omega }\tilde{l})
\right] +i\tilde{\beta}\left[  \left(  b^{2}+p_{\omega }^{2}\right) \sin(
p_{\omega}\tilde{l})\! -\!2p_{\omega}b\sin( b\tilde{l}) \right]
}{\tilde{l}p_{\omega}\left(  p_{\omega}^{2} -b^{2}\right)  \left(
\tilde{\omega}^{2}-b^{2}+i\alpha_{b}\tilde{\omega }\right)  \left[  \left(
1+\tilde{\beta}^{2}\right)  \sin(p_{\omega}\tilde
{l})+2i\tilde{\beta}\cos(p_{\omega}\tilde{l})\right]  }\right]  \nonumber
\end{align}
In the case of weak dissipation the total Josephson current can be split
into the dissipation and radiation parts, $I_{s}=I_{s,\mathrm{\
dis}}+I_{s,\mathrm{\ rad}}$. The radiation part plays the same role as
dissipation because in both cases energy is transfered from the moving
vortex lattice to other degrees of freedom (to photons in the case of
radiation). In the lowest order in $\lambda/\lambda _{J}\ll1$ at the
resonance frequencies we get
\begin{equation}
I_{s,\mathrm{\ dis}}\approx\frac{\alpha\varepsilon_{i}\omega_{p}
l}{2\varepsilon_{d}\omega d}I_{s,\mathrm{\ rad}}\approx\frac{\Phi_{0}
cw\alpha\tilde{l}\sin^{2}(b\tilde{l}/2)}{32\pi^{2}\lambda\lambda_{J}
b^{2}\tilde{\omega}|D|^{2}}.
\end{equation}
Losses due to radiation are equivalent to those caused by a resistor with
$Rw=2\pi/(\varepsilon_{d}\omega)$ attached parallel to JJ ($Rw\approx90$
ohm$\cdot$cm for $\varepsilon_{d}=1$ and $\nu=10$ GHz). The power fed into
JJ is $\mathcal{P}=IV$. Part of it, $(I_{t}+I_{s,\mathrm{\ dis}})V$, is
dissipated inside JJ, while another part, $I_{s,\mathrm{\ rad}}V$, is
radiated. Neglecting non-resonant part, $I_{t}$, we obtain for the
radiated fraction, $Q\equiv\mathcal{P}_{\mathrm{rad}}/\mathcal{P}$,
\begin{equation}
Q=\operatorname{Re}[\beta]\frac
{\tilde{\omega}}{2\tilde{l}}\frac{|\theta_{\omega}(0)|^{2}+|\theta_{\omega
}(-\tilde{l})|^{2}}{\alpha_{t}\tilde{\omega}+i_{J}}\label{RadFractionExact}
\end{equation}
In the regime of weak dissipation and near resonances we approximately
obtain
\begin{equation}
Q\approx \frac{r}{1+r},\ \ \ r=\frac{I_{s,\mathrm{\ rad}}}{I_{s,\mathrm{\
dis}}} =\frac{2\varepsilon_{d}d\omega}{\varepsilon_{i}\omega_{p} \alpha
l}. \label{RadFracApprox}
\end{equation}
To clarify the physical meaning of the parameter $r$, we can represent it
as $r=Q_{Z}(w\gg k_{\omega}^{-1})\mathcal{N}$ where
\[
Q_{Z}(w\gg k_{\omega}^{-1})=\frac{2\varepsilon_{d}d\omega}{\varepsilon_{i}\lambda_{J}\omega_{p}}\ll1
\]
is the transmission coefficient of the electromagnetic wave at the
junction edge (impedance mismatch) into free space in the limit $ w k_{\omega}\gg 1$
and $\mathcal{N}\!=\!1/\alpha\tilde{l}$
is the number of reflections before the Swihart wave decays inside JJ. At
$w k_{\omega}\sim 1$ and our result
is larger than that given by Eq.~(\ref{imp}) by the factor $\mathcal{N}$.
As dissipation inside JJ decreases ($\sigma_{t}$ and $\sigma_{q}$ drop),
$\mathcal{N}$ increases. As one can see from Eq.\ (\ref{RadFracApprox}),
the relation between the dissipative and radiative dampings is mainly
determined by competition between the two small parameters, $\alpha$ and
$d/l$. In the linear regime, $b\alpha\tilde{l}\tilde{\omega}\!>\!1$, we
get $\mathcal{N}<b\tilde{\omega}$. Due to limitation
$b<\lambda_{J}/\lambda$ we obtain
$Q\!\lesssim\!(d/\lambda)\tilde{\omega}^{2}\sim10^{-2}(\omega/\omega_{p})^{2}$.

\begin{figure*}[ptb]
\begin{center}
\includegraphics[width=6in,clip]{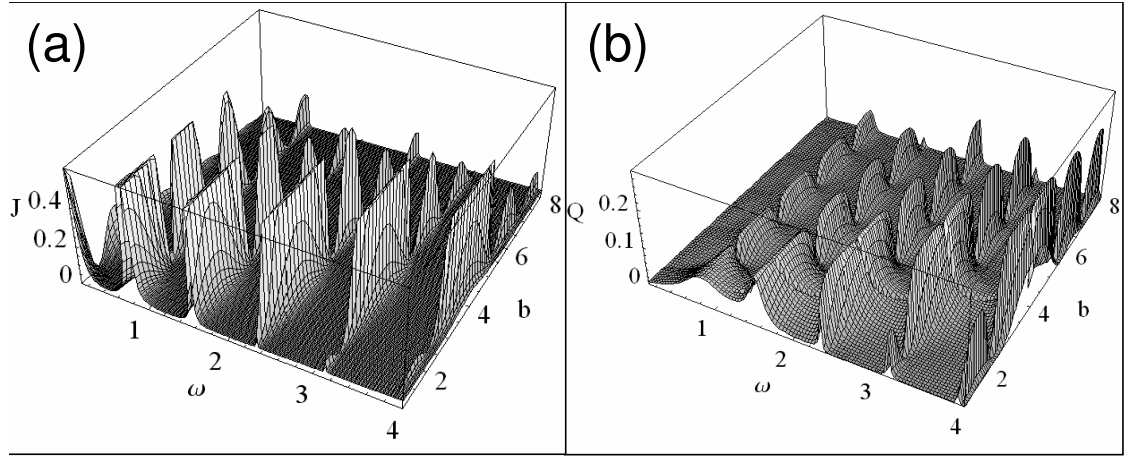}
\end{center}
\caption{Three-dimensional plots of the current density (left plot) and
the radiated fraction (right plot) for $\tilde{l}=4$, $\alpha_{t}=0.01$,
$\alpha_{q}=0$, and $\beta_{0} =0.001$ obtained within linear
approximation.} \label{J_Q3Dplots}
\end{figure*}

\begin{figure}[ptbptb]
\begin{center}
\includegraphics[width=0.5\textwidth,clip]{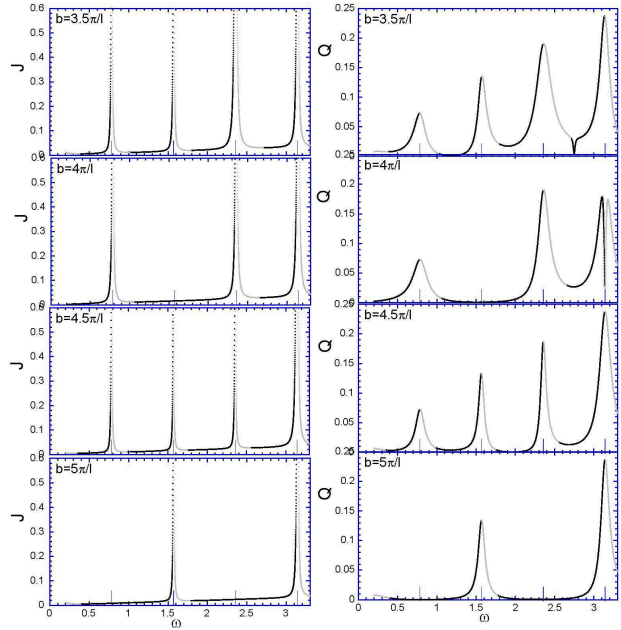}
\end{center}
\caption{The voltage dependencies of $J$ and $Q$ at several fields for the
same parameters as in the previous plot. The current shows sharp increases
near the marked Fiske-resonance frequencies. The linear approximation does
not describe narrow regions near the resonances. $Q$ reaches maxima given
by Eq.\ (\ref{RadFracApprox}) at the resonances. In the plots for
$b=3.5\pi/l\approx2.75$ and $4\pi/l\approx3.53$ one can see sharp drops of
$Q$ at the frequencies, corresponding to the Eck resonance condition
$\tilde{\omega}=b$.} \label{at0_01eta0_001}
\end{figure}

\begin{figure}[ptbptb]
\begin{center}
\includegraphics[width=0.5\textwidth,clip]{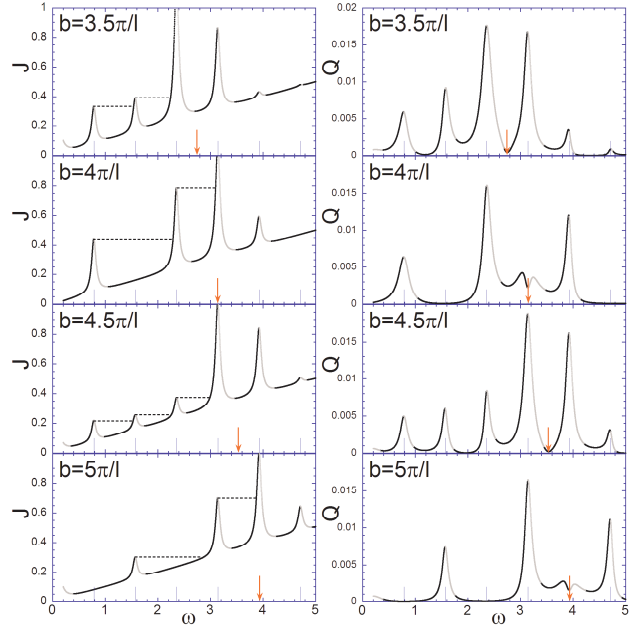}
\end{center}
\caption{The voltage dependencies of current $J$ and radiated fraction,
$Q$, at several fields for the higher dissipation $\alpha_{t}=0.1$ (other
parameters are the same as in the previous plots). Both $J$ and  $Q$ show
maxima near the marked resonance frequencies. Arrow marks frequency
corresponding to the Eck resonance condition $\omega =b$. With current
sweep the system will switch between different peaks, as it is illustrated
by the dashed lines.} \label{at0_1eta0_001}
\end{figure}

Figure \ref{J_Q3Dplots} shows three-dimensional plots illustrating
dependencies of the total reduced current density $J/J_{0}=\alpha_{t}
\tilde{\omega}+i_{J}$ and the radiated fraction $Q$ on the Josephson
frequency $\omega$ (voltage) and field $b $. For illustration purposes we
used toy parameters, $\tilde{l}=4$, $\alpha_{t}=0.01,$ $\alpha_{q}=0$, and
$\beta _{0}=0.001$. Figure \ref{at0_01eta0_001} shows voltage dependencies
of $J$ and $Q$ for several fields. The current shows well-known sharp
peaks at the Fiske-resonance frequencies, and peak amplitudes oscillate
with the magnetic field \cite{Kulik,Cirillo,Chang}. The conversion
coefficient  $Q$ shows smooth oscillating behavior reaching maxima given
by estimate (\ref{RadFracApprox}) at the resonances. We also observe
another nontrivial feature, $Q$ sharply drops at the position
corresponding to the Eck resonance conditions, $\tilde{\omega}=b$, even
though no feature is seen in $I$-$V$ dependencies for this frequency. For
comparison we also plotted in Fig.\ \ref{at0_1eta0_001} the same
dependencies for the case of higher dissipation, $\alpha_t=0.1$. This case
is quantitatively described by the linear approximation and allows us to
trace switching between different Fiske peaks with increasing current.

For a JJ with closed and open edges we obtain
\begin{align}
&  \theta_{\omega}(0)\approx-\sin(p_{\omega}\tilde{l})/(\tilde{\omega
}bD),\nonumber\\
&  D\approx\cos(\tilde{\omega}\tilde{l})+(i/2)(\beta+\alpha\tilde{l}
)\sin(\tilde{\omega}\tilde{l}),
\end{align}
The Poynting vector of radiation from open edge at $x=0$ is given by the
expression
\begin{equation}
\mathcal{P}_{\mathrm{rad}}\approx\frac{\Phi_{0}^{2}\omega\omega_{p}^{2}
w}{64\pi^{3}c^{2}b^{2}}\frac{\sin^{2}(\tilde{\omega}\tilde{l})}{|D|^{2}},
\end{equation}
The $I$-$V$ characteristics defined by Eq.~(\ref{c}) consists of maxima at
the frequencies $\omega_{n}$. The part of it, where $dI/dV<0$, is unstable
in the current-biased regime. Experimentally only voltages corresponding
to the resonance frequencies $\omega_{n}$ were observed \cite{Dm,Lan}.
Numerical calculations \cite{Ust,Ped} show that such a behavior occurs
in the nonlinear regime, i.e. at very low dissipation.

\subsection{Conclusions for single-junction case}

We have shown that in the case of weakly dissipative junctions $Q$ may
become of order unity in the linear regime. Nevertheless, the radiation
power per unit width, $\mathcal{P}_{\mathrm{rad}}/w$, given by
Eq.~(\ref{Poynting}), is always small in this regime because \textit{the
condition} $|\theta_{\omega }|\!\lesssim\!1$ \textit{also restricts the
power fed into JJ}. The open question is whether it is possible to get
$|\theta_{\omega}|\gg1$ and larger $\mathcal{P}_{\mathrm{rad}}$ in
strongly nonlinear regime when dissipation is very low and many Swihart
modes are involved. The $I$-$V$ characteristics in this limit have the form
of sharp steps (FIske steps) according to experimental data \cite{Dm,Lan} and numerical
calculations \cite{Ust,Ped}. However, the amplitude $\varphi_{\omega}(0)$,
which determines the radiation power, in highly nonlinear regime was not calculated yet.

In conclusion, accounting for the radiation into the dielectric outside
media, we derived the dynamic boundary conditions for JJ with the width $w$
much larger than the wave length of the electromagnetic wave radiated into
the free space. The method of derivation may be extended to the case $wk_{\omega}\ll 1$,
and following qualitative conclusions are valid for this case as well.
We have shown that in the linear regime of Josephson oscillations the radiated
fraction $Q$ of the power fed into the junction is determined by the
number of multiple reflections (i.e. by dissipation rate inside JJ) and by
the transmission coefficient, $Q_{Z}$, from JJ into free space. Even if
$Q$ reaches unity, radiation power per unit width of JJ remains small in
the linear regime. To probe radiation power from JJ in the nonlinear
regime, numerical study based on Eqs.~(\ref{st}) and (\ref{t}) is needed.
We think that measurements of radiation in the best junctions available
now, like those studied in Ref.~\onlinecite{Mart}, at low temperatures and
at intermediate magnetic fields may show higher radiation than that
observed previously.

\section{Radiation in layered superconductors}

For layered superconductors we need to formulate equations for the phase
differences $\varphi_n(t,x)$ inside each intrinsic junction $n$ between
layers $n$ and $n+1$ as well as the boundary conditions for these
variables. Here coordinates of layers are $y=(n+1/2)s$ and layers are
parallel to the plane $(x,z)$. In general $\varphi_n$ depend also on the
coordinate $z$, the direction of the applied magnetic field $H_0$, but we
will neglect this dependence (straight-vortex approach) assuming that
width of the crystal along the $z$-axis is much larger than the radiation
wavelength, $k_{\omega}w\gg 1$. The derivation of equations for
$\varphi_n(t,x)$ \cite{Bul,Kleinereq,Bul2} is similar to that for a single
JJ. Using the Maxwell equations, the expression for the intralayer
supercurrents via the phase of the superconducting order parameter inside
layers and the Josephson relation for interlayer current one need to
exclude all variables describing intralayer currents and electromagnetic
fields induced by these currents by expressing them via $\varphi_n$.
Formulation of the boundary conditions is also similar to a
single-junction case and is based on Eq.~(\ref{tot}). However, solution of
the equations for phase differences is now much more complicated because
the vortex structure is two-dimensional (along $x$ and $y$ axis) and may
vary depending on the parameters such as the applied dc magnetic field
$H_0$, the length $l$ along the $x$-axis and also on the transport current
(velocity of moving lattice). To have significant radiation power motion
of vortex lattice in different intrinsic junction should be synchronized,
as in multiple artificial JJ. For that interaction of intrinsic junctions
should be strong enough and it should favor in-phase vortices in all
junctions, i.e rectangular vortex lattice. First, we present the equations
and the boundary conditions for $\varphi_n(t,x)$ and then discuss
solutions for these equations and corresponding radiation power in linear
regime of Josephson oscillations when perturbation theory may be used to
solve equations for the phase differences. We will focus on crystals with
large number of layers on the scale of the London penetration length
$\lambda_{ab}$ for intralayer currents. In this case super-radiation from
many layers becomes possible when there are many junctions on the scale of
radiation wavelength and when vortices in many junctions are synchronized.
We will show that in this case significant part of the energy fed into the
crystal is converted into the radiation.

\subsection{Equations for the phase differences}

\begin{figure}[ptb]
\begin{center}
\epsfig{file=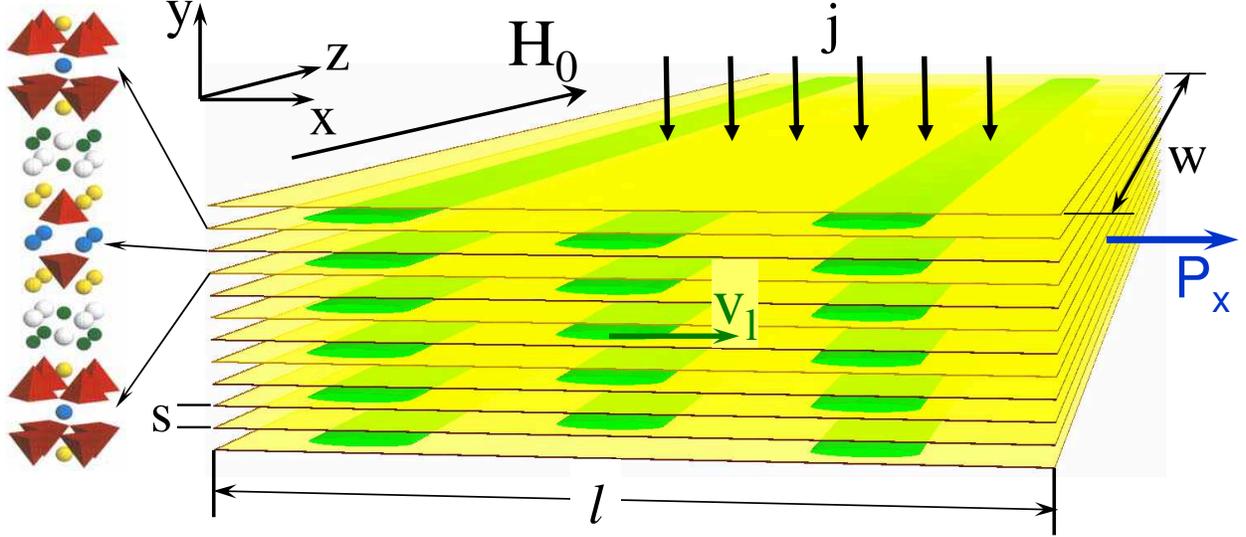,width=1.0\textwidth,angle=0}
\end{center}
\caption{Schematic picture of layered structure of superconductor and
Josephson vortex lattice. The unit cell of
Bi$_{2}$Sr$_{2}$CaCu$_{2}$O$_{8}$ compound and the relation of the crystal
structure to the layered model are shown in the left. The directions of
the applied dc magnetic field $H_{0}$, of the dc transport current $j$,
and of the radiation Poynting vector $P_{x}$ are also shown. The Josephson
vortices forming triangular lattice are shown schematically by elliptic
cylinders and $v_l$ is the velocity of moving lattice.} \label{Fig3}
\end{figure}

The crystal length $l$ is along the $x$-axis, while the transport current
is perpendicular to the layers. Thus, the Josephson vortex lattice moves
along the $x$ axis, see Fig.~\ref{Fig3}.

As we assumed that there is no dependence of the phase difference, current
density $\mathbf{j}$, and the electromagnetic fields on $z$ coordinate,
the components $j_{z}$, $E_{z}$, $B_{x}$, and $B_{y}$ vanish. The phase
difference $\varphi_{n}(t,x)$ between the layers $n$ and $n+1$ is defined
as
\begin{equation}
\varphi_{n}(t,x)=\Phi_{n}-\Phi_{n+1}-\frac{2\pi}{\Phi_{0}}\int_{ns}
^{(n+1)s}dyA_{y}(x),
\end{equation}
where $\Phi_{n}$ is the phase of the superconducting order parameter
inside the layer. Intrinsic junction $n$ is in the space
$(n-1/2)s<y<(n+1/2)s$. If all $N$ intrinsic Josephson junctions are
identical then the dynamics of the system can be described by
reduced coupled equations for the phase
differences, $\varphi_{n}(t,x)$, and reduced magnetic fields $h_{n}
=B_{y}(y=ns)/B_{c}$ with $B_{c}\equiv\Phi_{0}/(2\pi\lambda_{ab}\lambda_{c}
)$ (see. e.g., Ref. \cite{ak})
\begin{align}
\frac{\partial^{2}\varphi_{n}}{\partial\tau^{2}}+\nu_{c}\frac{\partial
\varphi_{n}}{\partial\tau}+\sin\varphi_{n}-\frac{\partial
h_{n}}{\partial u}
&  =0,\label{LayEqPhase}\\
\left(  \nabla_{n}^{2}-\ell^{-2}\left(
1+\nu_{ab}\frac{\partial}{\partial \tau}\right)  \right)
h_{n}+\left(  1+\nu_{ab}\frac{\partial}{\partial\tau
}\right)  \frac{\partial\varphi_{n}}{\partial u}  &  =0. \label{LayEqField}
\end{align}
Here we used reduced $x$ coordinate, $u=x/\lambda_{J}$ normalized to the
Josephson length $\lambda_{J}=\gamma s$ and reduced time,
$\tau=\omega_{p}t$, where $\omega_{p}=c/(\lambda_{c}\sqrt{\epsilon_{c}})$
is the plasma frequency, $\epsilon_{c}$ is the $c$-axis high-frequency dielectric
constant inside the superconductor, $\lambda_{ab}$ and $\lambda_{c}$ are
the London penetration lengths, $\ell\equiv\lambda_{ab}/s$, $s$ is the
interlayer distance, and $\nabla _{n}^{2}$ notates the discrete second
derivative operator, $\nabla_{n} ^{2}A_{n}=A_{n+1}+A_{n-1}-2A_{n}$. The
dissipation parameters, $\nu_{ab}
=4\pi\sigma_{ab}/(\gamma^{2}\epsilon_{c}\omega_{p})$ and $\nu_{c}=4\pi
\sigma_{c}/(\epsilon_{c}\omega_{p})$, are determined to the quasiparticle
conductivities, $\sigma_{ab}$ and $\sigma_{c}$, along and perpendicular to
the layers, respectively. The conductivity $\sigma_{ab}$ plays the same
role as the conductivity $\sigma_{q}$ for a single JJ. Applying the
operator $\nabla_{n}^{2}-\ell^{-2}\left(
1+\nu_{ab}\frac{\partial}{\partial\tau }\right)  $ to the first equation
and excluding $h_{n}$, we can also derive
equations containing only $\varphi_{n}(u,\tau)$\thinspace
\begin{equation}
\left[  \nabla_{n}^{2}-\ell^{-2}\left( 1+\nu_{ab}\frac{\partial}{\partial
\tau}\right)  \right]  \left(  \hat{T}_{c}\frac{\partial\varphi_{n}}
{\partial\tau}+\sin\varphi_{n}\right)  +\left( 1+\nu_{ab}\frac{\partial
}{\partial\tau}\right) \frac{\partial^{2}\varphi_{n}}{\partial u^{2}}=0,
\label{bulk}
\end{equation}
with $\hat{T}_{c}\equiv\partial/\partial\tau+\nu_{c}$. Eqs.~(\ref{bulk})
represent just charge conservation laws. The first term describes the
change of electron charges inside the layers and Cooper-pair interlayer
tunneling currents, while the second term describes superconducting
intralayer currents. The terms with the coefficient $\nu_{ab}$ describe
quasiparticle dissipative in-plane currents induced by moving Josephson
vortices. The static interaction of junctions is described by the term
$\nabla_{n}^2\sin\varphi_n$, while their dynamic interaction is described
by the term $T_c(\partial/\partial\tau)\nabla_{n}^2\varphi_n$. Both are
short-range (nearest-neighbor) weak interactions and they are not very
effective in keeping long-range ordering of vortex lattice along the
$y$-axis.

In the system of finite number of layers $N$ the dc current with the
density $J$ is injected in layer 1 and extracted from layer $N$. Then
equations for the first and the last junction are obtained from
Eq.~(\ref{bulk}) by putting $\varphi_{0}=\varphi_{N+1}=0$ in linear terms
and replacing $\sin\varphi _{0}=\sin\varphi_{N+1}=j$. In a finite-layer
system the edge junctions differ strongly from other junctions because
they have only one neighboring junction.

The electric field inside the superconductor between layers $n$ and
$n+1$ in terms of the phase difference is given as
\begin{equation}
E_{y;n,n+1}\approx\frac{\Phi_{0}}{2\pi
sc}\frac{\partial\varphi_{n}}{\partial
t}. \label{e}
\end{equation}
The average (over time or space) electric field determines the
Josephson frequency $\omega_{J}$. The average magnetic field we
denote by $B$ and we introduce dimensionless average magnetic field
$b=2\pi s\lambda_{J}B/\Phi_{0}$.

The parameters of BSCCO at low temperatures are $\omega_{p}/(2\pi)=0.15$
THz, the Josephson critical current $J_{c}=1700$ A/cm$^{2}$, $\gamma=500$,
$\epsilon_{c}=12$, $s=15.6$ \AA , $\sigma _{c}(0)=2\cdot10^{-3}$
(ohm$\cdot$ cm)$^{-1}$, $\sigma_{ab}(0)=4\cdot10^{4}$ (ohm$\cdot$
cm)$^{-1}$ \cite{LatKB} and so $\ell\approx 100$, $\nu_{ab} \approx0.2$
and $\nu_{c}\approx5\cdot10^{-4}$.
An important feature of the high-temperature superconductors is higher
relative in-plane dissipation in comparison with $c$ axis dissipation,
$\nu_{ab}\gg\nu_c$. Another layered high-T$_c$ compound
Tl$_2$Ba$_2$CaCu$_2$O$_8$ has lower anisotropy $\gamma\approx 150$ and as
a consequence higher critical current $J_{c}\approx 3\cdot10^{4}$
A/cm$^{2}$ and higher plasma frequency, $\omega_{p}/(2\pi )\approx0.75$
THz \cite{Thor}.

\subsection{Boundary conditions and the Poynting vector of radiation}

We assume that there are only outcoming waves and use Eq.~(\ref{tot}) for
an open edge.  In the junction $n$ we approximate $B_z(y)\approx B_z(y_n)$
and $E_y(y)\approx E_y(y_n)$, where $y_n=sn$. This gives the following
boundary condition at the boundary $x=0$,
\begin{eqnarray}
&&B_z(\omega, n)=\sum_mU(\omega,n-m)E_y(\omega,m), \label{bclayered}\\
&&U(\omega,n)\approx (1/2)s[|k_{\omega}|J_0(k_{\omega}sn)+ik_{\omega}N_0(k_{\omega}sn)],
 \ \ n\neq 0,\nonumber
\end{eqnarray}
while for $n=0$ we need to substitute $-(2/\pi)\ln [1/(|k_{\omega}|s)]$
for $N_0(k_{\omega}sn)$. Finally, we use the relations
\begin{equation}
B_z(t,n)=B_c\ell^2\left[\frac{\partial\varphi_n}{\partial u}\right]_{x=0},
\ \
E_y(t,n)=\frac{B_c\ell}{\sqrt{\epsilon_c}}\left[\frac{\partial\varphi_n}{\partial
\tau}\right]_{x=0}, \ \ B_c=\frac{\Phi_0}{2\pi\lambda_{ab}\lambda_c}
\end{equation}
to formulate the boundary condition for oscillating part of
$\varphi_n(x,t)$. This gives the following general boundary conditions for
the phase differences
\begin{equation}
\frac{\partial\varphi_{n}}{\partial u}=\pm\frac{is\tilde{\omega}}{2\ell\sqrt
{\varepsilon_{c}}}\sum_{m}[|k_{\omega}|J_{0}(k_{\omega}s|n-m|)+ik_{\omega
}N_{0}(k_{\omega}s|n-m|)]\varphi_{m}
\label{BoundCondLayGen}
\end{equation}
In the time representation we obtain again Eq.~(\ref{space-time}), as for a single
junction.

For the Poynting vector we derive
\begin{equation}
P_x(\omega)=\frac{\Phi_0^2\omega^3}{64\pi^3c^2N s}\sum_{n,m}
J_0(k_{\omega}s|n-m|)\varphi_n(\omega,0)\varphi_m^*(\omega,0).
\label{poylay}
\end{equation}

\begin{figure}
\begin{center}
 \epsfig{file=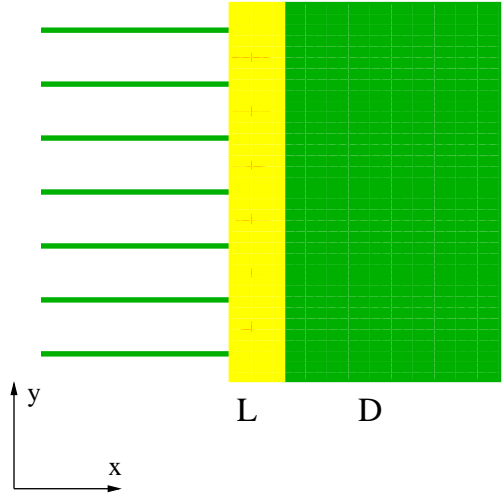,width=0.4\textwidth,angle=0}
\end{center}
\caption{ The stack of intrinsic Josephson junctions at $x<0$, dielectric
at $0<x<L\ll k_{\omega}^{-1}$ and a superconducting screen with the London
penetration length $\lambda_s$ and the thickness $D\gg \lambda_s$. Such a
screen introduces additional interaction of vortices in different
intrinsic junctions.} \label{Fig7}
\end{figure}

Let us consider now the layered superconductor with a superconducting screen
at the right side separated from the layered crystal by the dielectric
with thickness $L$ small in comparison with the wavelength of the
electromagnetic wave in this dielectric, see Fig.~\ref{Fig7}.

Then, as in the case of a single JJ,  we neglect effect of the dielectric
on the boundary conditions. The solution inside the superconducting screen
is given by Eq.~(\ref{scr}). From this solution we obtain
\begin{equation}
B_z(\omega,y)=\frac{i\lambda_{s,\omega}^{-2}}{k_{\omega}}\int dy'K_0\left(\frac{y-y'}{\lambda_{s,\omega}}\right)E_y(\omega,y').
\end{equation}
This relation leads to the boundary condition for the oscillating part of the phase
difference
\begin{equation}
\frac{\partial\varphi_n(\omega,0)}{\partial u}=\frac{\lambda_Js}{\lambda_{s,\omega}^2}
\sum_mK_0\left[\frac{s(n-m)}{\lambda_{s,\omega}}\right]\varphi_m(\omega,0).
\label{bcscreen}
\end{equation}
Here space derivative of the phase differences and the phase differences are
at the point $x\rightarrow -0$. The space derivative now is not small, the
parameter $(s\lambda_J/\lambda_s^2)$ may be of order unity or larger. The
term in the free energy, ${\cal F}_{bc}$, corresponding to the boundary
condition Eq.~(\ref{bcscreen}), is
\begin{eqnarray}
&&\frac{{\cal F}_{bc}}{w}=\int dy\int dx\frac{B_z^2}{4\pi}=
\lambda_{s,\omega} s\sum_n\frac{B_z^2(sn)}{4\pi}= \label{bc4} \\
&&=\frac{\Phi_0^2s}{16\pi^3\lambda_{s,\omega}^3}
\sum_{n,m}H\left[\frac{s(n-m)}{\lambda_{s,\omega}}\right]
\varphi_n(\omega,0)\varphi_m^*(\omega,0), \ \ \
H(an)=\sum_kK_0(ak)K_0[a(k+n)]. \nonumber
\end{eqnarray}
This free energy can be compared with the bulk inductive interaction, corresponding to
Eq.~(\ref{bulk}), see Ref.~\onlinecite{kb},
\begin{equation}
{\cal F}_{{\rm ind}}=\frac{\Phi_0^2}{32\pi^3\lambda_{ab}\lambda_c}\int du\nabla_u
\varphi_n(u)\nabla_u\varphi_m(u)\exp\left[-\frac{|n-m|s}{\lambda_{ab}}\right].
\label{ind}\end{equation}
Both interactions favor triangular lattice at least in the static case.
The coefficient in Eq.~(\ref{bc4}) is much larger than that in the
inductive free energy, i.e. superconducting screen enhances strongly the
tendency to form triangular lattice along the $y$-axis for not very large
junction length $l$.

\subsection{Solutions for the phase difference and the radiation power at
high fields in large-$N$ case}

We consider here the simplest case of layered crystals with large number
of layers $N\gtrsim \ell$,
when we can neglect edge effects along the $y$-axis, i.e.
the difference between the edge (the first and the last) and inner junctions.
This corresponds to the
thickness larger than $0.2$ $\mu$m with the total number of junctions $N>100$.

In the linear regime of Josephson oscillations the general solution for
the phase difference has the form
\begin{equation}
\varphi_n(\tau,u)=\tilde{\omega}\tau-bu+\kappa_n+\theta_n(\tau,u),
 \ \ \theta_n(\tau,u)\ll 1.
\end{equation}
For $\kappa_n=0$ the lattice is rectangular, $\varphi_n(\tau,u)$ are
$n$-independent. For $\kappa_n=\pi n$ the lattice is triangular, vortices
in neighboring layers are in anti-phase positions. For a static lattice
such a configuration minimizes the energy of the magnetic field inside the
crystal with large $l,N$. However, at small $l$ boundary conditions are
inconsistent with triangular lattice at some values of $bl$, and in this
case rectangular lattice becomes more favorable \cite{KoshPhysC06}. For
moving vortex lattice energy consideration does not work, and here the
parameters $\kappa_n$ should be determined by the condition that total
current $I$ via each junction is the same \cite{ak},
\begin{equation}
J_c\lambda_Jw\int_{-\tilde{l}}^0du\langle\sin\varphi_n(\tau,u)\rangle_{\tau}=I,
\ \ .
\end{equation}
where $J_c=c\Phi_0/(8\pi^2s\lambda_c^2)$ is the Josephson current
density.  Slowly moving lattice preserves its triangular structure.
This was confirmed experimentally via observation of magnetic
oscillations of the flux-flow resistivity with the period of one
flux quantum per two junctions \cite{Hirata}. Theoretical analysis
\cite{ak,ArtPRB03} shows that the triangular lattice becomes
unstable at the lattice velocity slightly smaller than the Swihart
velocity $c_S=cs/(2\lambda_{ab}\sqrt{\epsilon})$. This instability
corresponds to experimentally observed end point of the first
flux-flow branch. It was also shown that interaction with top and
bottom surfaces leads to significant lattice deformations \cite{ak}.

Situation at high velocity $v_l\gg c_S$ is less clear. Lattice structures
and their stability in this regime in the case of large lateral size $l$
have been addressed in Ref.\ \cite{ak} and have been reconsidered in Ref.
\cite{ArtPRB03} with the conclusion that for parameters typical for BSCCO
no stable regular lattice exists at high velocities. Structures and
stability of steady states for small lateral sizes $l$ is an open issue.

In the following sections we will estimate the radiation power for
rectangular and triangular lattice and for lattice with random values of
$\kappa_n$. Those are most probable realizations of vortex configurations
in the large $N$ limit. They give also an estimate for maximum radiation
power which one can anticipate.

\subsection{Rectangular vortex lattice}

\begin{figure}
\begin{center}
 \epsfig{file=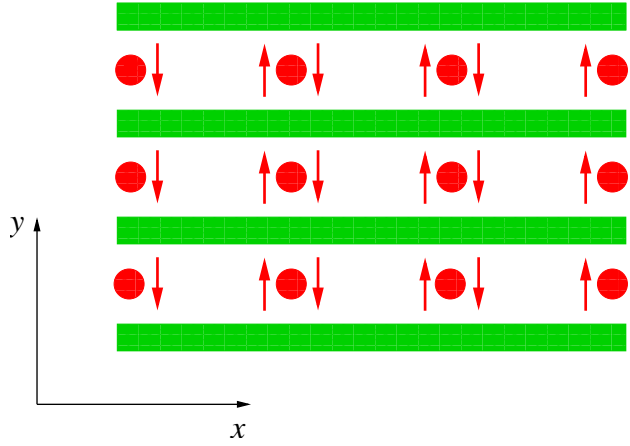,width=0.5\textwidth,angle=0}
\end{center}
\caption{Rectangular lattice of Josephson vortices in layered
superconductors. The screening currents flow along the $y$ axis leading to
the large energy of the magnetic field (inductive coupling).
Correspondingly, the amplitude of phase modulation is weak, but vortices
come to the junction edges coherently inducing super-radiation.  }
\label{Fig8}
\end{figure}

For rectangular lattice sketched at Fig.~\ref{Fig8},
$\varphi_n=\tilde{\omega}\tau-bu+\theta(\tau,u)$, the equation for the
oscillating part, $\theta(u)$, is given by
\begin{equation}
\frac{\partial^2\theta}{\partial\tau^2}+
\nu_c\frac{\partial\theta}{\partial\tau}-\ell^2\frac{\partial^2\theta}{\partial
u^2} \approx -\sin(\tilde{\omega}\tau-bu). \label{req}
\end{equation}
For a junction opened at both edges the solution is similar to that of a
single JJ, Eq.~(\ref{aa}). However, there is important difference due to
the presence of large coefficient $\ell^2$ in front of the second space
derivative. This coefficient is due to uniformity of rectangular lattice
and corresponding supercurrents along the $y$-axis, see Fig.~\ref{Fig8}.
This uniformity leads to large energy of the magnetic field, i.e.
inductive coupling, Eq.~(\ref{ind}), {\it resulting in small amplitude of
phase variations}.
To find radiation power and $I$-$V$ characteristics we use results of the
perturbation theory for a single JJ. To make Eq.~(\ref{req}) similar to
that for a single JJ we introduce $\tilde{u}=u/\ell$, $\tilde{b}=b\ell$
and $\tilde{l}$ should be replaced by $\tilde{l}/\ell$. We limit
 ourself to the frequencies much smaller than the Eck resonance
$\tilde{\omega}\ll b\ell $. In this case the solution has the form
$\theta(\tau,\tilde{u})= \operatorname{Re}\left [ \theta_{\omega}\exp(-i \omega
\tau)\right]$ with
\begin{equation}
\theta_{\omega}(u)\approx
e^{i\tilde{b}\tilde{u}}/(i\tilde{b}^{2})+a_{1}\exp(ip_{\omega}\tilde{u})
+a_{2}\exp(-ip_{\omega}\tilde{u}), \label{aa1}
\end{equation}
The coefficients $a_{1}$ and $a_{2}$ should be found from the boundary
conditions given by Eq.~(\ref{BoundCondLayGen}). For rectangular lattice at
the edge $x=0$ we obtain
\begin{equation}
\frac{\partial\theta_{\omega}(0)}{\partial \tilde{u}}=
\frac{i\tilde{\omega}}{2\sqrt{\epsilon_c}}s\sum_n\left[|k_{\omega}|J_0(k_{\omega}sn)+
ik_{\omega}N_0(k_{\omega}sn)\right]\theta_{\omega}(0).
\label{BoundCondRectGen}
\end{equation}

\subsubsection{Large number of junctions $N>(k_{\omega} s)^{-1}$}

We consider first the case of very large number of layers $N>(k_{\omega} s)^{-1}$. Then
\begin{equation}
\frac{\partial\theta_{\omega}(0)}{\partial \tilde{u}}
\approx
\frac{i\tilde{\omega}}{\sqrt{\epsilon_c}}\theta_{\omega}(0).
\end{equation}
This corresponds to $\beta\approx 1/\sqrt{\epsilon_c}$. Important point is
that for such $\beta$ {\it the perturbation theory is correct for
$\tilde{\omega}|b\ell-\tilde{\omega}|\gg 1$, i.e. practically at all
interesting fields and frequencies}. We limit ourself to the frequencies
smaller than the Eck resonance $\tilde{\omega}\ll b$. Thus the amplitude
of Swihart waves is proportional to small factor $\ell^{-1}$ and
oscillating part of the phase differences at $Nsk_{\omega}\gtrsim 1$ is
described by Eqs.~(\ref{solsing}) with the extra factor $\ell^{-1}$ for
$\theta_{\omega}(0)$ and $\theta_{\omega}(-l)$. Further, we need to
replace $\tilde{\omega}\tilde{l}$
 by $\tilde{\omega}\tilde{l}/\ell\approx 0$, while in $D$ we need to put
$\beta=1/\sqrt{\epsilon_c}$ and $\alpha=\nu_c$. Assuming that the length
of junctions is small enough, $\tilde{\omega}\tilde{l}\ll\ell$ and using
results Eq.~(\ref{solsing}) we see that oscillatory dependence of the
radiation power and of the dc current on $\tilde{\omega}\tilde{l}$ drops
out and we obtain for the dc current at $\tilde{\omega}b\ell\gg 1$ and
$\tilde{\omega}\tilde{l}\ll\ell$ the expression {\it without resonances}
\begin{equation}
i=\frac{I}{J_cwl}=\nu_c\tilde{\omega}+
\frac{\sqrt{\epsilon_c}|\sin(b\tilde{l}/2)|} {b^2\ell\tilde{l}}
\frac{1}{\tilde{\omega}}. \label{iv}
\end{equation}
For the radiation power of $N$ layers, using Eq.~(\ref{poylay}), we obtain
\begin{equation}
\frac{\cal{P}_{{\rm rad}}}{w} \approx \frac{\Phi_{0}^{2}
\omega_{p}^{2}\epsilon_c}{32\pi^{3}cb^{2}}
\frac{N}{s\ell^2}\sin^2(b\tilde{l}/2). \label{genepowerlay}
\end{equation}
Comparing this result with that for a single junction,
Eq.~(\ref{genepower}), we see that the additional factor
$N/(k_{\omega}s\ell^2)$ is present. In BSCCO and TBCCO crystals at the
frequency $\nu=1$ THz the parameter $k_{\omega}s\ell^2\sim 1$. Thus the
radiation  power of $N$ layers for a moving rectangular lattice is
$N/(k_{\omega} s)$ times larger than from a single intrinsic junction and
additional large factor $1/(k_{\omega} s)$ is due to super-radiation in
the case $Nk_{\omega}s\gtrsim 1$.

Let us discuss now the $I$-$V$ characteristics given by Eq.~(\ref{iv}).
At a given current $i>i_{{\rm min}}$ we have stable solution with positive slope,
$dI/dV>0$, at $\tilde{\omega}>\tilde{\omega}_{{\rm min}}$.
Here
\begin{equation}
\tilde{\omega}_{{\rm min}}=
\left(\frac{\sqrt{\epsilon_c}}{b^2\ell\tilde{l}\nu_c}\right)^{1/2}
|\sin(b\tilde{l}/2)|, \ \ i_{{\rm min}}=2\nu_c\omega_{{\rm min}}.
\end{equation}
Hence, moving rectangular lattice cannot exist at currents $i<i_{{\rm
min}}$ due to super-radiation as power fed into the crystal should support
radiation as well as quasiparticle dissipation needed to ensure stability
of this dynamic state. This condition is necessary but it may be not sufficient
for the stability of moving rectangular lattice. The ratio of radiation power
to that of dissipation one is $r=(\tilde{\omega}_{{\rm
min}}/\tilde{\omega})^2$. Thus maximum value of $Q$ is 0.5 at
$i=i_{{min}}$, and $Q$ drops as current increases beyond $i_{{\rm min}}$.

For BSCCO crystals at $\tilde{l}=\pi$ and $b=1$ we estimate
$\tilde{\omega}_{{\min}}\approx 4$, while $i_{{\min}}\approx 4\times
10^{-3}$. At the frequency $\nu=1$ THz we obtain ${\cal P}_{{\rm
rad}}/w\sim N$ ($\mu$W/cm) for Tl$_2$Ba$_2$CaCu$_2$O$_8$ crystals. Using
the energy conservation law, $IV={\cal P}_{{\rm rad}}$, we see that to
reach this power the total dc current $I$ via the junction should be about
0.004 of the critical current. The radiation power in BSCCO crystals at
the same frequency is about $25$ times weaker due to smaller Josephson
plasma frequency.

\subsubsection{Moderate number of junction $\ell<N\ll(k_{\omega}s)^{-1}$}

Consider now practically more interesting case of moderate number of
layers, $\ell<N\ll(k_{\omega}s)^{-1}$. In this case from Eq.\
(\ref{BoundCondRectGen}) we obtain the following boundary condition
\begin{align}
\frac{\partial\theta_{\omega}(0)}{\partial u} &
=\frac{i\tilde{\omega}\beta_{N}}{\ell}\theta_{\omega}(0),\\
\beta_{N} &  =\frac{sN}{2\sqrt{\epsilon_{c}}}
\left[|k_{\omega}|-i\frac{2k_{\omega}}{\pi}\ln\frac{1}{|k_{\omega}|sN}\right].
\end{align}
This condition is very similar to the boundary condition
(\ref{BoundSmoothFinal}) for a single junction.

Solving for $a_1$ and $a_2$, we obtain for the edge phase
\begin{equation}
\theta_{\omega}(0)\approx\frac{\cos(\tilde{\omega}\tilde{l}/\ell)-\exp(
-ib\tilde{l}) }{\ell b\tilde{\omega}\left[  \sin(\tilde{\omega}
\tilde{l}/\ell)+i(  2\beta_{N}+\nu_{c}\tilde{l}/2\ell)
\cos(\tilde{\omega}\tilde{l}/\ell)\right]  }
\end{equation}
Due to $\beta_N\ll 1$ at small dissipation we obtain the resonant-type
$I$-$V$ characteristics with Fiske resonances at
$\tilde{\omega}\tilde{l}/\ell=\pi n$, as in a single JJ.
Moreover, for parameters of BSCCO the c-axis quasiparticle dissipation is
negligible in comparison with the radiation damping. At resonances we
derive
\begin{equation}
i_{J}   =\nu_c\tilde{\omega}+\frac
{1-(-1)^{n}\cos(b\tilde{l})}{2\tilde{l}\tilde{\omega}\ell b^{2}{\rm
Re}[\beta_{N}]},
\end{equation}
The linear approximation brakes down at $2\ell b\tilde{\omega}{\rm
Re}\beta_{N}\sim1$, i.e., it is valid for number of layers $N
>\sqrt{\epsilon_{c}}/(b^2\tilde{\omega}\lambda_{ab}|k_{\omega}|)$. At the
frequency 1THz  linear approximation is valid if $N>100$. The total
radiation power per unit width at the resonance frequency \emph{does not depend on
number of layers} and is given by
\begin{equation}
\frac{{\cal P}_{\rm rad}}{w}
=\frac{\epsilon_{c}\Phi_{0}^{2}\omega_{p}^{2}}{32\pi
^{3}\lambda_{ab}^{2}\omega b^{2}}\left[1-(-1)^{n}\cos(b\tilde{l}) \right].
\label{RadRectSmallN}
\end{equation}
The independence of this result on $N$ appears as a result of cancellation
of the $N^2$ factor in the total power of $N$ coherently radiating
junctions and the factor $\beta_N^{-2}\propto N^{-2}$ which determines the
resonance damping. Away from resonances and for resonances limited by
quasiparticle damping ${\cal P}_{\rm rad}\propto N^{2}$. For BSCCO
crystals at $\omega/2\pi=1$ THz, we obtain ${\cal P}_{{\rm rad}}/w\approx
24$ mW/cm, while for Tl$_2$Ba$_2$CaCu$_2$O$_8$ crystals we estimate ${\cal
P}_{{\rm rad}}/w\approx 0.5$ W/cm. Now $\tilde{\omega}_{{\rm min}}$ and
$i_{{\rm min}}$ become larger as $N$ drops:
\begin{equation}
\tilde{\omega}_{{\rm min}}^{3/2}=
\left(\frac{\sqrt{\epsilon_c}}{b^2\ell\tilde{l}\nu_c}\right)^{1/2}
\frac{2|\sin(b\tilde{l}/2)|}{\sqrt{\omega_psN/c}}, \ \ i_{{\rm
min}}=\frac{3}{2}\nu_c\omega_{{\rm min}}.
\end{equation}
As $N$ increases beyond $N\sim \ell$ the total radiation power  does not
depend on $N$ in Fiske resonances and increases $\propto N^2$ outside
resonances, while $\tilde{\omega}_{{\rm min}}$ decrease as $N^{-1/3}$.
This behavior holds until $N$ reaches $(k_{\omega}s)^{-1}$. After that
${\cal P}_{{\rm rad}}$ increases linearly with $N$, while
$\tilde{\omega}_{{\rm min}}$ remains constant.

It is interesting to compare our estimate with recent large-scale
simulations of THz radiation out of BSCCO mesa by Tachiki {\it et al.}
\cite{Tach}. They solve the Maxwell equations coupled with the equations
for the intralayer phases inside the crystal and the dielectric media.
They found that close to one of the Fiske resonances lattice is mostly
disordered but with pronounced rectangular correlations promoted by
generated standing electromagnetic wave. They found quite powerful outside
radiation in this state, with power density up to $P_x=3000$ W/cm$^2$. Our
estimate following from Eq.\ (\ref{RadRectSmallN}) for 100 junctions
occurs to be only slightly smaller $\sim$ 1500 W/cm$^2$.

\subsection{Triangular lattice}

Triangular lattice, again in the limit $b\gg 1$, behaves quite differently
in comparison with rectangular one. The solution has the form
\begin{equation}
\varphi_n(\tau,u)=\tilde{\omega}\tau-bu+\pi n+(-1)^n\phi(\tau,u)+
\theta(\tau,u). \label{tri}
\end{equation}
Here $\phi$ describes the amplitude of the phase oscillations with
Josephson frequency $\omega$, while $\theta$ describes oscillations with
the frequency $2\omega$ which are induced by moving vortex lattice due to
nonlinearity of coupled sine-Gordon equations (\ref{bulk}), see
Ref.~\onlinecite{Art}.
According to theoretical estimates \cite{ak} when the lattice velocity
approaches the Swihart velocity, the lattice may generate a very powerful
electromagnetic wave \emph{inside superconductor}, with power density up
to $20$W/cm$^2$. Unfortunately, this main harmonic of electromagnetic wave
at the frequency $\omega$ experiences full internal reflection at the
boundary and does not radiate outside. The triangular lattice produce only
weak outside radiation at frequency $2\omega$, caused by the homogeneous
in c direction component of the phase, $\theta(\tau,u)$. In the case of
semiinfinite superconductor the radiation power at $2\omega $ has been
estimated in by Artemenko and Remizov \cite{Art}. Here we estimate this
power for finite-size samples when pronounced Fiske resonances are
present.

Putting the solution Eq.~(\ref{tri}) into Eq.~(\ref{bulk}) we obtain
coupled equations for $\phi(\tau,u)$ and $\theta(\tau,u)$:
\begin{eqnarray}
&&\frac{\partial^2\phi}{\partial\tau^2}+
\frac{\partial}{\partial\tau}\left(\nu_c-\frac{\nu_{ab}}{4} \frac{\partial^2}{\partial u^2}\right)\phi+
 \frac{1}{4} \frac{\partial^2\phi}{\partial u^2}
=-\sin(\tilde{\omega}\tau-bu), \\
&&\frac{\partial^2\theta}{\partial\tau^2}+ \nu_c\frac{\partial
\theta}{\partial\tau}- \ell^2\frac{\partial^2\theta}{\partial u^2}=
-\phi\cos(\tilde{\omega}\tau-bu),
\end{eqnarray}
From the second equation we see that $\theta$ is of order $\ell^{-1}$ and,
as a consequence,  terms linear in $\theta$ were omitted in the first
equation in comparison with the term describing dissipation.
\begin{figure}
\begin{center}
 \epsfig{file=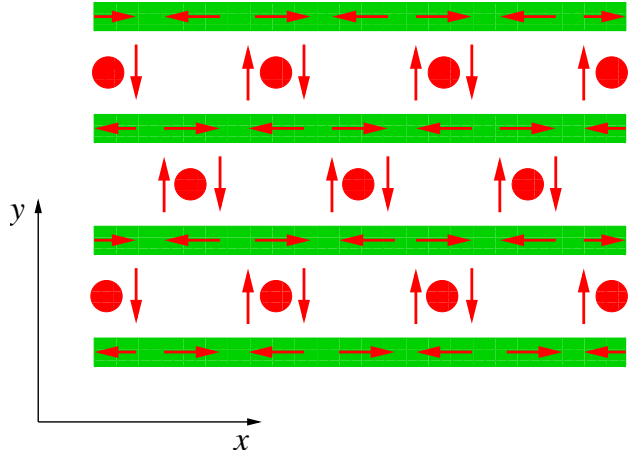,width=0.5\textwidth,angle=0}
\end{center}
\caption{Triangular lattice of Josephson vortices in layered
superconductors. Screening currents compensate each other inside the
lattice vortex cell ($2s$,$1/b$) leading to significant reduction of the
magnetic energy and enhancement of the amplitude of the phase variations
in comparison with rectangular lattice. The in-plane screening currents
lead to intralayer quasiparticles contribution to the dissipation of the
moving lattice. Due to nonlinearity there are field components at the
frequency $2\omega$ with small amplitude. They cause super-radiation at
the frequency $2\omega$.}
 \label{Fig9}
\end{figure}

The boundary conditions for the alternating and homogeneous phase are
given by
\begin{equation}
\frac{\partial\phi(0)}{\partial u}=0, \ \ \
\frac{\partial\theta(0)}{\partial u}\approx \pm
i\beta\tilde{\omega}\theta(0).
\end{equation}
for $u=0,\tilde{l}$ with $\beta=1/\sqrt{\epsilon_c}$. In the framework of
the perturbation theory, at $\tilde{\omega}b\alpha_{{\rm
tr}}\tilde{l}\gtrsim 1$, the amplitude $\phi$, in the lowest order in
$1/b$, is given by the expression
\begin{equation}
\phi= \frac{4[ \cos(q(u+l))-\exp(  -ibl) \cos(qu)] } {bq\sin(ql) }, \ \
q=2\tilde{\omega}+\alpha_{{\rm tr}}, \label{phiK}\end{equation} where
$\alpha_{{\rm tr}}=\nu_c+\nu_{ab}\tilde{\omega}^2/4$. Putting it into the
equation for $\theta$ and accounting for boundary conditions we obtain
\begin{equation}
\theta\approx\frac{i}{2\tilde{\omega}\beta\ell b^{3}}[  1-\exp(-2ib\tilde{l})]
-\frac{ [\cos(2\tilde{\omega}\tilde{ l})-\exp( -ib\tilde{l}) ] [
1-\exp( -ib\tilde{l})]  }{\omega^{2}\beta\ell b^{2}[
\sin(2\tilde{\omega}{ l})+i\alpha_{\mathrm{tr}}\tilde{l}\cos(2\tilde{\omega}\tilde{l})]}.
\end{equation}
Keeping only the resonance term, we obtain the radiation power at the
frequency $2\omega$ in the linear regime
\begin{equation}
{\cal P}_{{\rm
rad}}(2\omega)=\approx\frac{\varepsilon_{c}\Phi_{0}^{2}\omega_{p}^{4}}{2\pi^{3}
cs^{2}\ell^{2}\omega^{2}b^{4}} \frac{N}{k_{\omega}s}
 \frac{[ \cos^{2}(2\tilde{\omega}\tilde{ l})-2\cos(2\tilde{\omega}\tilde{ l})
\cos (b\tilde{l}+1][  1-\cos(  b\tilde{l}) ]  }{
\sin^{2}(2\tilde{\omega}\tilde{ l})+(  \alpha_{\mathrm{tr}}\tilde{l})  ^{2}\cos^{2}(
2\tilde{\omega}\tilde{ l})}.
\end{equation}
Here, as for rectangular lattice, we have large factor $(k_{\omega}s)^{-1}$ due to super-radiation at the frequency $2\omega$.
This differs from the radiation power for rectangular lattice at the same
frequency by the factor of order unity at $\tilde{\omega}\sim 1$ and $b\sim 1$ in the linear regime.
Thus both rectangular and triangular lattice give approximately
the same radiation power for $\tilde{\omega}\sim 1$ and in both cases the
radiation is coherent. However, their $I$-$V$ characteristics and the
conversion coefficients $Q$ are very different.
The fraction
of the power converted into radiation depends on quasiparticle
conductivities $\sigma_{ab}$ and $\sigma_c$. These conductivities remain
nonzero in cuprate superconductors even if temperature approaches zero
because cuprates are gapless superconductors. As a result, only small part, of
order $(b^2\ell^2\alpha_{{\rm tr}}\tilde{l})^{-1}$, is converted into the
radiation in the linear regime. Thus for triangular lattice we estimate
very small $Q\propto \ell^{-2}$. This translates into strong heating and
practically triangular lattice cannot provide a realistic source of radiation.

The $I$-$V$ characteristics is determined by the behavior of $\phi$ and so
it is resonant at low dissipation, as in a single JJ. In particular, for
(i) weak dissipation, $q\approx 2\tilde{\omega}+i\alpha_{\mathrm{tr}}$,
and (ii) small frequencies $\tilde{\omega}\ll b/2$ we obtain
\begin{equation}
i_{J}\approx\frac{8}{b^{4}}\alpha_{b}\tilde{\omega}
+\frac{2\alpha_{\mathrm{b}}\left[ \cos(2\tilde{\omega}\tilde{
l})-\cos(b\tilde{l})\right] \cos\left( 2\tilde{\omega}\tilde{ l}\right)
}{\tilde{l}\tilde{\omega} b^{2}\left[ \sin^{2}(2\tilde{\omega}\tilde{ l})
+\left(  \alpha_{\mathrm{tr}}l\right)  ^{2}
\cos^{2}\left(  2\tilde{\omega}\tilde{ l}\right)  \right]  },
\end{equation}
where $\alpha_b=\nu_c+\nu_{ab}b^2$ and $\ \alpha_{\mathrm{tr}}=\nu_{c}
+\nu_{ab}\tilde{\omega}^{2}$. The amplitude of the Fiske peak at
$2\tilde{\omega}_{n}\tilde{l}=\pi n$ is
\begin{equation}
\delta_{n}i_{J}\approx\frac{2\left[  1-(-1)^{n}\cos(b\tilde{l})\right]
}{\tilde{\omega} _{n}b^{2}\tilde{l}^{3}\left(
\nu_{c}+\nu_{ab}\tilde{\omega}_{n}^{2}\right)  }.
\end{equation}
Note that triangular lattice excites resonances for the antiphase modes
whose frequencies are by factor $2\ell=2\lambda{ab}/s\sim 300$ smaller
then the frequencies of the homogeneous modes excited by the rectangular
lattice. Very pronounced Fiske steps have been observed recently by Kim
{\emph et al.} \cite{Kim04}

We also estimated that the lattice with random parameters $\kappa_n$ gives
incoherent radiation with approximately the same power and the conversion
coefficient as triangular lattice.

\subsection{Conclusions}

We have estimated the radiation power from most probable vortex lattices
in the large-$N$ limit. Our estimate for rectangular lattice in BSCCO
agrees satisfactory with results of numerical study by Tachiki {\it et
al.} \cite{Tach}. We see that only rectangular lattice gives significant
radiation power and quite high conversion coefficient, i.e. up to half
power fed into the crystal may be converted into the radiation. Thus, not
very high currents are needed to get significant radiation power. However,
the main question to be answered is what are parameters of the stability
of rectangular lattice in finite-size samples. Other open questions
include:
\begin{enumerate}
\item What is the radiation power in highly nonlinear regime for
triangular lattice and for rectangular lattice with moderate number of
layers $N$ ?%
\item What is the structure of steady states and the radiation power of
crystals with not very large $N<\ell$, when the boundaries along the
$y$-axis become important?%
\item What are boundary conditions and the radiation power in crystals
with $wk_{\omega}\ll 1$ ? %
\item Are there  \emph{fully gapped} layered superconducting materials or
artificial multilayer system which can be used as radiation sources?
Presence of finite gap in the whole Fermi surface drastically reduces
quasiparticle dissipation at low temperatures and
the problem of high dissipation for the moving triangular lattice
in the high-temperature superconductors will be avoided.
\item Is it possible to get stronger radiation from the triangular lattice
by modulating the radiated edge of the crystal? For example, one can
use some periodic layer of dielectric or metal between the crystal and the
air which may result in more effective conversion of high $k_y$
electromagnetic fields inside the crystal into low $k_y$ outside
electromagnetic waves. For that the surface modulation has to have the
periodicity $2s$, i.e., every second layer has to be closed. For efficient
conversion, one can anticipate that the radiation power would be larger by
the factor $\ell^2\approx 10^4$ in BSCCO and TBCCO crystals.
\end{enumerate}

The authors thank Yu. Galperin, A. Gurevich, I. Martin, K. Rasmussen, N.
Gr\o nbech-Jensen, and R. Kleiner for useful discussions. This research
was supported by the Department of Energy under contracts \# W-7405-ENG-36
(LANL) and \# W-31-109-ENG-38 (Argonne).

\end{document}